\documentclass{aa}
\usepackage{epsfig}
\usepackage{natbib}
\usepackage{lscape}
\begin{document}

\title{Search for Associations Containing Young stars (SACY)\subtitle{I. Sample \& Searching Method}
\thanks{Based on observations made under the ON-ESO agreement for the joint operation of the 1.52~m ESO telescope,
on two 2.2~m ESO telescope runs, 
and at the  Observat\'{o}rio do Pico dos Dias 
MCT/Laborat\'{o}rio Nacional de Astrof\'{\i}sica (LNA/MCT),  Brazil}$^,$
\thanks{Tables 3, 4 and 5 are only available in electronic form
at the CDS via anonymous ftp to cdsarc.u-strasbg.fr (130.79.128.5)
or via http://cdsweb.u-strasbg.fr/cgi-bin/qcat?J/A+A/}}

 \author{
        C. A. O. Torres\inst{1},
        G. R. Quast\inst{1},
        L. da Silva\inst{2},
        R. de la Reza\inst{2},
        C. H. F. Melo  \inst{3,4},
\and    M. Sterzik\inst{3}
   }

   \institute{
    Laborat\'{o}rio Nacional de Astrof\'{\i}sica/MCT, Rua Estados Unidos 154, 37504-364 Itajub\'{a}, Brazil
\and
    Observat\'{o}rio Nacional/MCT, 20921-400 Rio de Janeiro, Brazil
\and
        European Southern Observatory, Casilla 19001, Santiago 19, Chile
\and
    Departamento de Astronom\'{\i}a, Universidad de Chile, Casilla 36-D, Santiago, Chile
}

\date{Received / Accepted}

\abstract{

We report results from a high-resolution optical
spectroscopic survey aimed to search for  nearby young associations
and young stars among optical counterparts of ROSAT All-Sky Survey
X-ray sources in the Southern Hemisphere. We selected 1953
late-type ($B-V~\geq~0.6$), potentially young, optical counterparts out
of a total of 9574 1RXS sources for follow-up observations. At
least one high-resolution spectrum was obtained for each of 1511
targets. 
This paper is
the first in a series presenting the results of the SACY survey. Here we describe
our sample and our observations.  We describe a convergence method in the (UVW) velocity space 
to find associations. As an example, we discuss the validity of this
method in the framework of the $\beta$~Pic Association.

            }
  \authorrunning{C. A. O. Torres et al.}
  \titlerunning{SACY: Sample \& Searching Method}
\maketitle

\section{Introduction}

Post-T Tauri Stars (pTTS), and its prototypical example FK Ser \citep{Herbig73}, were introduced as a new,
systematic, category by \cite{Herbig1978} as T Tauri stars (TTS) that follow classical and weak line TTS (cTTS and wTTS) 
in an evolutionary sequence.
The pTTS were expected to outnumber the cTTS and wTTS populations by far, but their discovery and identification remained
difficult for a long time. 
The relation of pTTS to "isolated" TTS \citep{Quast1987} (young,
low-mass pre-main sequence stars found spatially far away from any apparent dark or parental molecular cloud) remained
an open problem. Interestingly, Herbig's catalog of pTTS, and the suggested list of isolated TTS produced independently
by the Brazilian group are actually quite similar.    An intriguing case is TW~Hya, a high Galactic latitude object,
confirmed as TTS \citep{RucinskiKrautter1983}, with a distance of at least  13$\degr$ from the nearest  dark clouds.
Its origin remained enigmatic for a long time.

A systematic search for more isolated TTS was pursued with the  optical spectroscopic 
Pico dos Dias Survey (PDS) among  optical counterparts of the 
IRAS Point Source Catalog  
\citep{gregorio92, Torres1995, Torres1998}.
One of the first results of the PDS
was the discovery of four additional TTS  around TW~Hya  \citep{delareza89, gregorio92}. 
They concluded that this group was likely a very young association relatively close to the sun.

The putative TW~Hya association was actually a very odd result.
In fact, in a recent review,  \cite{zuckermanSong04}  reminded us
that until the late 1990 the  Hyades and the UMa clusters 
(with an age about 600~Myr and 300~Myr, respectively)
were the only coeval, co-moving concentrations of stars known within 60~pc from Earth.
Thus, the closest place to find  very young stars (typically about 10~Myr) 
was in the nearby classical star forming regions, such as Sco-Cen, Taurus, $\rho$~Oph 
(at $\sim$150~pc) and Orion somewhat farther at $\sim$450~pc.

Later, \cite{Kastner97} confirmed that TW~Hya forms a physical association
of about 20~Myr and at a distance of 40 to 60~pc from Earth -- the TW~Hya Association (TWA),
based on the similarity of the X--ray fluxes, radial velocities, astrometry (Hipparcos) and their spectroscopic characteristics. Even additional members were found later \citep {webb99, sterzik99, zuckerman01, song03}.

Only very few good candidates for isolated TTS were found with the PDS.
But while the IR-excess selection criteria  effectively finds young stellar objects 
embedded in their placental material or with circumstellar disks, 
it fails to signal older objects whose disks have been dissipated. 
Hence most stars  with ages between $\ga10-70$~Myr (i.e. wTTS and pTTS)  
escaped the discovery by this method.

Due to the enhanced X-ray activity of young stars \citep{Walter86} more efficient selection criteria for post
and isolated TTS candidates were developed. The high sensitivity and full sky coverage of the ROSAT all-sky survey
\citep{truemper} revealed thousands of new X--rays sources projected in the direction of nearby star
forming regions \citep{Guillout94}. Ground-based spectroscopic studies showed that a large fraction
of these X--rays sources were indeed wTTS  together with older pTTS and ZAMS stars. 
\citep[e.g.][]{Alcala00}. Surprisingly, many of the newly found weak-line TTS were {\it not} obviously connected to any molecular cloud
region which again raised many questions about their origin \citep{Sterzik95}.

Young, nearby associations, and in particular TWA, are of great importance to understand the local star forming history, and they reveal samples of stars that allow to study the transition phase of disk dispersal and planet formation in great detail. Therefore several groups have started to 
look for similar nearby young associations hidden among the ROSAT X--ray sources. 
Shortly after the discovery of TWA eight more young nearby associations have been announced
\citep[see \cite{torres03} and the review by][]{zuckermanSong04},
namely,  $\eta$~Cha \citep[$d\sim97$~pc and age $\sim9$~Myr;][]{mamajek99, Lyo};
 $\epsilon$~Cha \citep [$d\sim115$~pc and age $\sim5$~Myr;][] {mamajek00, feige2003};
the Horologium Association (HorA) \citep[$d\sim60$~pc; age $\sim30$~Myr;][]{torres00}
and the adjacent and very similar Tucana Association (TucA) simultaneously discovered by \cite{zucketal01};
the $\beta$~Pic Association (BPA)  \citep[$d\sim35$~pc; age $\sim15$~Myr;][]{barrado98, zuckerman01};
the AB~Dor Association \citep[$d\sim20$~pc; age $\sim50$~Myr;][ - calling it as AnA]{zuckerman04, torres03};
and finally two less well defined associations,
the Argus Association \citep[$d\sim100$~pc; age $\sim30$~Myr;][]{torres03} and
the Octans Association \citep[$d\sim110$~pc; age $\sim30$~Myr;][]{torres03}.

The similarity between the adjacent TucA and HorA motivated us to start SACY (Search for Associations Containing Young
stars)\footnote{Sacy is a mythological Brazilian being, a little black boy with only one leg, that frightens night
travelers.} survey in 2000. 
Some early obsevations, obtained in 1999 for other purposes, were also included in the SACY, whose
initial goal was to examine the physical relation between these two associations. 
Later, SACY was enlarged in order to look for other, hitherto undiscovered, nearby young associations among the optical
counterparts of the ROSAT X--ray sources.

Preliminary results of the SACY  have been presented in different conference
contributions \citep[][and references therein]{torres03}, where we have suggested
eleven potential nearby associations 
using the preliminary data (including components of the Sco-Cen Association).
One of those first results was the
existence of two somewhat superposed associations which we called GAYA1 
and GAYA2\footnote{GAYA stands for Great Austral Young Association.}.
GAYA1 is so similar to the  Tucana/Horologium Association proposed
by \cite{zuckermanSong04} that we can adopt this designation, although some of their proposed members seem
to match better with GAYA2 ($d\sim80$~pc; age $\sim20$~Myr), as we will present in a forthcoming paper.
In fact, the present paper is the first one of a series that aims to present the results of the SACY.
Here, we present the survey, the observed sample and the observing strategy (Sec. 2).
In the associated catalog, we give the spectroscopic properties (radial velocity, $V\sin(i)$, spectral type,
Lithium 6708~\AA\ and $H_\alpha$ equivalent widths) 
as well as the \mbox{$UBV(RI)_C$} photometry  of the observed stars.
Our kinematical approach is presented and exemplified with the known BPA \citep{zuckerman01}
for which we propose new members.

\section{The sample and observations}

\begin{figure*}
\begin{center}
  \resizebox{0.8\hsize}{!}{\includegraphics[bb=24 250 550 780,clip]{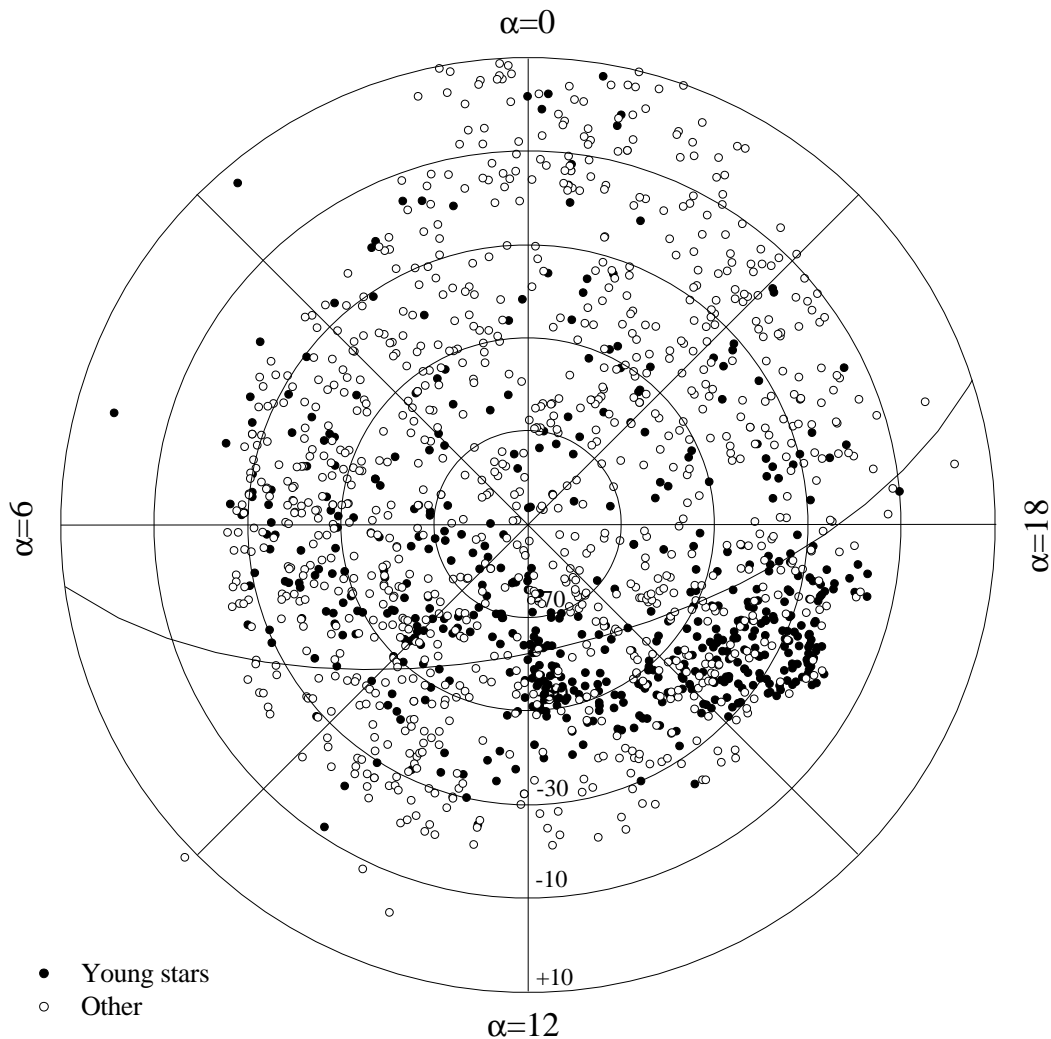}}
\caption{Celestial polar projection of the stars observed in SACY.
Stars classified as young according to their Li\,I equivalent width (see section 4.2) are plotted as
filled circles. More evolved stars are indicated as open circles.  The transverse curve represents the Galactic plane.)}
\label{fig:pizza}
\end{center}
\end{figure*}

The ROSAT All-Sky Bright Source Catalogue (1RXS) \citep{Voges99} was used as a starting point to build 
an initial sample containing young star candidates. 
Therefore, 1RXS was cross-correlated with the Hipparcos and
Tycho-2 Catalogues. 
We followed   an approach  similar to that described by \cite{Guillout99}. 
Within an error radius of 2.6 times the ROSAT positional error one (or more) possible optical
counterparts were selected for each 1RXS source. Only stars having $B-V\geq0.6$ (from Hipparcos or Tycho-2) were
considered. This cut-off is justified  by the lithium depletion behavior. For young stars with spectral types later
than mid-G (i.e., $B-V\sim0.7$) the strength of the Li 6708~\AA\ line can be used as an youth indicator
\citep{martin97}. Our cut-off (0.6) corresponds approximately to a G0 dwarf, in order not to loose interesting objects
close to this boundary. We exclude hotter stars, because they lack of a similar youth indicator and due to the
difficulty to obtain precise  radial velocities $(V_r)$ necessary for the convergence method later employed in SACY.

We also excluded all Hipparcos stars having $Mv~<~2.0$, because for our color limit young  stars older than 5~Myr must
be less luminous than $10L_{\odot}$. There is some subjectivity in our final list, as the positions, velocities,
parallaxes, colors, and magnitudes are all affected by different errors, and we had to decide on a case by case basis
about inclusion in our final list. For example, some late F stars were included and some early G may have been
excluded, but these cases are not very important as we use a conservative color limit.   

\begin{table}
\caption{Completeness of SACY. The table gives the  northern boundary declination as a function of right ascension  up to which the SACY is complete.}
\label{table:comp}
\centering
\begin{tabular}{cc}
\hline
RA interval &   Declination\\
(h)     &   $(\degr)$ \\
\hline
\hline
$01  -  05$  & $-20$ \\
$05  -  09$  & $-25$ \\
$09  -  16$  & $-20$ \\
$16  -  18$  & $-15$ \\
$18  -  22$  & $-10$ \\
$22  -  01$  & $+10$ \\
\hline
\end{tabular}
\end{table}

From an initial list containing 9574 1RXS sources in the Southern Hemisphere, we found 1953  
of them having counterparts in the Tycho-2  and Hipparcos Catalogues using the above criteria. 
During the last seven years, 1511 stars
were observed at least once. Data from  the literature for 115  
objects were used to complement the observed sample.
(Other 153 
objects were observed that do not fulfill the SACY criteria.)

In Figure~\ref{fig:pizza} we show the observed sample and in Table~\ref{table:comp} we give the
completeness of SACY as a function of right ascension. 
 Only 3 selected sources were not observed within this area and there are also 
29 stars observed outside this area, 
mainly near the limits defined in the table. 
Some of the holes seen in Figure~\ref{fig:pizza} may be due to incompleteness of 1RXS.
The concentration of young stars between  RA~$\sim12$~h to ~$\sim17$~h, corresponding 
to the Sco-Cen Association, is clearly visible in Figure~\ref{fig:pizza}.

\subsection{The rejected sample}
Whenever available, SIMBAD information was used 
to exclude stellar counterparts of the 1953 southern sources being well known RS~CVn, W~UMa,
giant stars or any other contamination showing {\it ab initio} 
clear signs that they are not  young late type stars.
Within the area defined in Table~\ref{table:comp}
we rejected 51 sources and they are presented in Table~\ref{table:reject}.

In such way, we have covered $\sim80\%$ of the 1953 sources. 

\begin{table}
\caption{The rejected sample.
Column 1: ROSAT Identification; Column 2: Identification from Hipparcos or Tycho-2 Catalogues; 
Column 3: Spectral type or other reason for rejection; Column 4: Other identification.
}
{\scriptsize
\label{table:reject}
\centering
\begin{tabular}{clll}
\hline
1RXS & Ident & Type & Other \\
~~~(1) & ~(2) & (3) & (4)~  \\
\hline
\hline

J005307.5-743903   &Hip  4157  &RS        &CF Tuc  \\
J010945.2-201300   &Hip  5452  &WUMa      &CT Cet  \\
J011143.2-255720   &T6425-1882 &K0III+    &BZ Scl  \\
J014852.7-205340   &Hip  8447  &WUMa      &TW Cet  \\
J015951.5-380725   &Hip  9336 &K2III      &HD 12320 \\
J023401.2-653634   &Hip 11934  &WUMa      &WY Hor\\
J024208.5-463121   &Hip 12611 &K1III      &HD 17006 \\
J024326.2-375540   &Hip 12716  &RS       &UX For \\
J035941.3-584028   &Hip 18659 &K0III    &UY Ret\\
J040729.6-523413   &Hip 19248  &RS      &AG Dor         \\
J041425.3-204921   &T5889-1407 &K1III SB2  &HD 26917\\
J044145.7-471540   &Hip 21843 &G8III      &HD 30042\\
J050135.3-444955   &Hip 23381 &K0III      &HD 32517\\
J051833.2-681328   &T9162-0347  &WUMa      &RW Dor\\
J082401.6-694149   &T9198-0008   &A2 &HD311345\\
J085943.2-274850   &Hip 44164  &RS        &TY Pyx\\
J091756.4-372350   &Hip 45623 &G3III      &HD 80332\\
J100352.1-561134   &T8603-1735 &K1III      &HD 87525\\
J103733.8-564757   &T8609-1385   &XB        &4U1036-56\\
J110952.1-762909   &T9410-2627   &A7V       &CD-75 522 \\
J111510.2-611540   &Hip 54948   &Cl       &NGC3603\\
J112545.4-763035   &T9411-1577 &K0III      &HD 99558\\
J112557.8-401547   &T7743-1091 &G6II       &HD 99409\\
J113616.1-380210   &T7740-0698  &RS      &V858 Cen\\
J113922.1-392315   &Hip 56851  &RS      &V829 Cen\\
J114929.5-504849   &T8228-0126 &K1III      &HD102726\\  
J125850.0-365827   &Hip 63347  &WUMa    &V839 Cen\\
J132711.7-245126   &T6717-0230 &Gal Sey &\\
J135155.7-363726   &Hip 67682  &WUMa    &V757 Cen\\
J140710.6-510131   &T8276-0897 &Gal     &\\
J160546.1-395032   &T7855-0905 &D2"(A5V+?) &A:HD144118\\
J162804.7-491150   &T8320-2046   &XB   &4U1624-49\\
J163528.5-480556   &T8329-1465   &B8&HD330993\\
J172446.8-341224   &T7383-0007  &Cl&Pr 24\\
J172509.3-341117   &T7383-0162  &Cl&Pr 24  \\
J173728.0-290759   &T6839-0472 &Gal Sey    &\\
J175620.0-721850   &T9297-0956  &F2IV D0.8" &HD162128\\
J180108.7-250444   &T6846-0633   &XB        &4U1758-25\\
J180847.7-260037   &T6847-3167  &B9&HD315209\\
J181806.3-121439   &Hip 89681  &Cl        &NGC 6604 \\
J181842.3-134706   &T5689-1381  &Cl&NGC 6611  \\
J182030.8-161022   &T6265-1255  &Cl       &NGC 6618\\
J182615.1-145034   &T5702-1197   &XB O7V  &V4138 Sgr\\
J192240.0-203840   &Hip 95244 &K2III   &V4138 Sgr\\
J202052.5-300213   &T7442-0726 &Gal Cl  &ACO 3674\\
J213949.1-160018   &Hip 106961  &RS      &AD Cap\\
J214906.4-304157   &T7488-0390 &Gal Cl  &ACO 3814\\
J220036.3-024433   &Hip 108644 &G5III+sdOB &FF Aqr \\
J231324.1+024028   &Hip 114639 &K1III      &SZ Psc\\
J232748.5+045126   &Hip 115819  &WUMa    &VZ Psc\\
J234351.0-151655   &Hip 117054  &M4III     &R Aqr\\

\hline
\end{tabular}
}
\end{table}

\subsection{Spectroscopic observations}

Most of the spectroscopic observations ($\sim70\%$) were performed with the high-resolution 
($R\sim50000$)  FEROS \'echelle spectrograph  \citep{kaufer99}
at the 1.5~m/ESO  telescope at La Silla (Chile) between January 1999 and September 2002 
(ON-ESO agreement and  ESO program identification 67.C-0123).
In October 2002, FEROS was moved to the 2.2~m telescope.
After that, two more runs (ESO program identifications 072.C-0393, 077.C-0138) were carried out aimed mainly 
to collect additional 
radial velocity points to resolve the orbits of the
previously identified spectroscopic binaries. 
We noticed no relevant changes in the zero-point and/or in the spectrograph
resolution (see below).
The FEROS observations were taken in the OS (Object-Sky) mode.
The reduction of the spectra was performed by the FEROS pipeline including flat-fielding, 
background subtraction, removal of cosmic rays, wavelength calibration and 
barycentric velocity correction, yielding as the final product a 1D re-binned spectrum.

Another set of data ($\sim30\%$) was collected at the coud\'e spectrograph attached to
1.60~m telescope at the Observat\'{o}rio do Pico dos Dias (OPD), LNA, Brazil. 
The  setup used was centered at 6500~\AA\ covering a spectral range of 450~\AA\ 
with a resolution of $R\sim9000$.
These observations were reduced with standard IRAF packages.

A few observations were collected using the two-fiber-fed high-resolution spectrograph 
CORALIE ($R\sim47000$) \citep{queloz00}
attached to the Swiss Euler Telescope, also at La Silla, Chile. 
As for FEROS, all observations were taken in the OBJ2 mode, i.e.,
one fiber  centered on the target star and the other fiber illuminated by the background sky.
Similarly to FEROS, CORALIE also has an on-line reduction system which processes 
the raw frames following the same steps as described
above for FEROS pipeline.

\subsection{Photometric data}

\mbox{$UBV(RI)_C$}  photometry for part of the sample was obtained using FOTRAP
\citep{jablonski94} at the 0.60~m Zeiss telescope of the OPD.
When a star was not observed photometrically by us we try to obtain some useful 
photometric data in the Hipparcos
and Tycho Catalogues or in the available literature in the SIMBAD.
For multiple stars, magnitudes and colors  were corrected in order to account 
for the presence of the companion. 
We will use in the paper preferentially the $(V-I)_C$ color.
If no observed value is available we use  one deduced from the $(B-V)_T$ when $V\leq10$
or otherwise from the spectral type.

\section{The SACY Catalog}

The SACY Catalog contains the objects studied in our survey. 
The tables are only available in electronic form.
The  Table 3 ({\sc The SACY sample}) contains  the data for 1626 stars,  
1511 observed by us and  115 objects with  information from the literature.  
In the Table 4 ({\sc Other observed stars}) we list 153 observed stars that do not fulfill the SACY criteria.
The structure of both tables is identical, and they contain astrometrical data, radial and rotational velocities,
photometric and spectroscopic data.  
Bibliographic sources used in the catalog are given in the Table 5 ({\sc The bibliographic sources}).

\section{Analysis}
The primary goal of the SACY is to derive space motions for
 all young 1RXS sources in the Southern Hemisphere.
Thus, our main effort was to obtain  reliable radial velocities from the spectroscopic observations.
Combined with Hipparcos parallaxes and proper motions, they
allow us to compute the (UVW) components of the Galactic space velocity vector which, in turn,
is the key ingredient to find young associations (see below).
The spectroscopic observations also enable us to obtain the spectral types and  to measure both
the Li 6708~\AA\  and the $H_\alpha$ equivalent widths, which are used as youth indicators.

\subsection{Radial and rotational velocities}

Radial velocities were derived by cross-correlating our FEROS and  coud\'e  spectra with the spectrum
of the  radial velocity standard Tau~Cet (HD~10700)
observed at the same night and used as template.
During the periods where Tau~Cet was not visible, 3 other radial velocity standards
from the CORALIE Extra-Solar planet Survey were kindly provided by the Swiss Team and used as reference.
In very few nights no template could be observed.
In these unfortunate cases, the template observed in an adjacent night was used.
As we have not used very red templates, our radial velocities for red dwarfs may be less precise.

In order to spot eventual errors,  radial velocities were computed by two of us independently
using  the task {\it fxcor} from IRAF and the method described in \cite{melo01}.
This latter makes use of  K0 CORAVEL-type binary mask \citep{baranne79} as cross-correlation template.
In both methods, the resulting cross-correlation function (CCF) can be approximated by a
Gaussian function whose center readily gives the radial velocity
and the width is related to the broadening mechanism such as
turbulent motions, gravity pressure and rotation.

The photon noise errors on the radial velocity measurements computed as described
in \cite{melo01} are very low thanks to our good $S/N$ which is always better than 30-50.
Thus the uncertainties on the radial velocities are dominated by the spectrograph drift
during the night due to changes in the air index and atmospheric pressure.
These shifts are of the order of a few hundreds m s$^{-1}$ per night.
A more realistic idea of our overall precision can be obtained
by measuring the dispersion of the radial velocity standard Tau~Cet
measured almost every  night during our observing runs.
The mean radial velocity and its r.m.s are respectively -16.316~km~s$^{-1}$
and 0.320~km~s$^{-1}$ spanning over 1050 days of interval.

\begin{figure}[th!]
  \resizebox{\hsize}{!}{\includegraphics{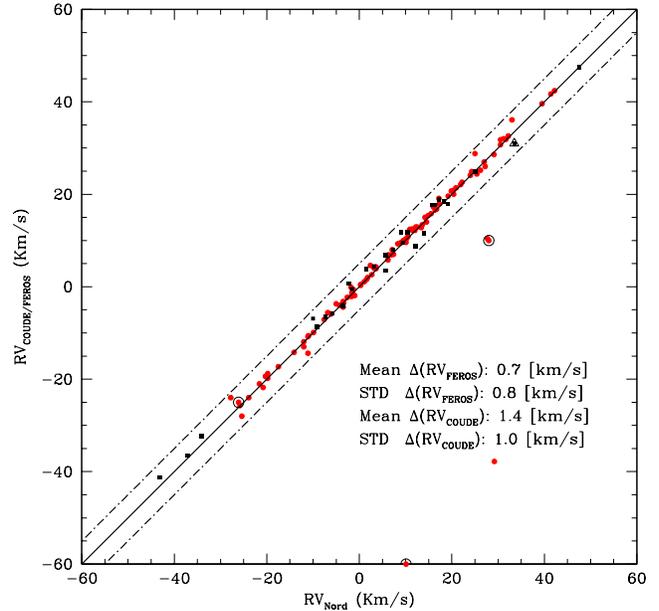}}
  \resizebox{\hsize}{!}{\includegraphics{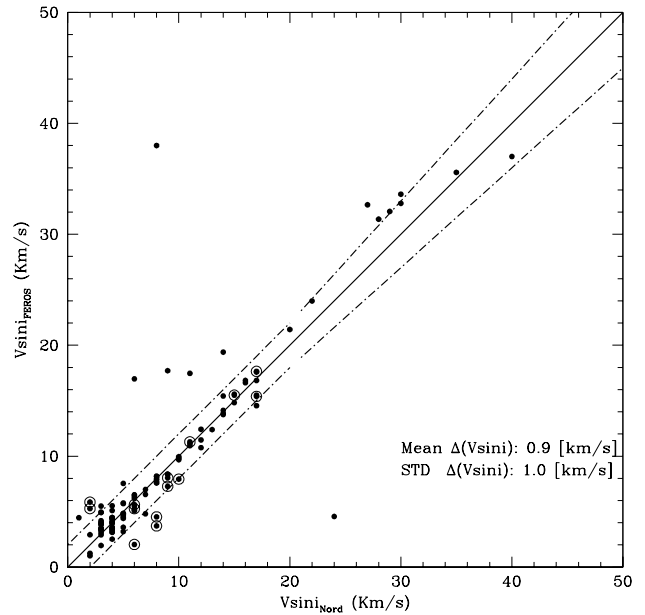}}
\caption{Comparison of our derived radial velocities and $V\sin(i)$ with those derived by 
\cite{Nord04}. {\it Top.}
Radial velocities obtained with FEROS are shown as filled circles whereas those 
obtained with the coud\'e are seen as filled squares.
The continuous line is the one-to-one relation and the two dashed-dotted lines delimit 
a $\pm$5~km~s$^{-1}$ region around the continuous line.
{\it Bottom.}  Again, the one-to-one relation is shown as a continuous line 
whereas the dashed-dotted lines mark the $V\sin(i)$ errors of 2~km~s$^{-1}$ for $V\sin~i\le20$~km~s$^{-1}$ 
and 10\% of the observed $V\sin(i)$ for $V\sin~i>20$~km~s$^{-1}$. 
Multiple stars are shown as encircled symbols.}
\label{fig:compnord}
\end{figure}

In Figure~\ref{fig:compnord} (top) we compare our radial velocity measurements with those from \cite{Nord04}.
Radial velocities obtained with FEROS are shown as filled circles
whereas those obtained with the coud\'e are seen as filled squares.
The continuous line is the one-to-one relation and the two dashed-dotted lines
delimit a $\pm$5~km~s$^{-1}$ region around the continuous line.
We have used the mean radial velocity error ($\epsilon(V_r)$) and the dispersion of the
radial velocity measurements ($\sigma(V_r)$) given by \cite{Nord04} to
exclude as much as possible bad quality radial velocity measurements
or multiple systems out of Figure~\ref{fig:compnord}.
In this sense, objects with  $\epsilon(V_r)>1$~km~s$^{-1}$ or
with $\sigma(V_r)>3\times\epsilon(V_r)$ were excluded.
The remaining known binaries not excluded by these criteria are overploted by circles and triangles,
respectively for FEROS and coud\'e observations.
Thus, excluding the known binaries and {\it one} point whose $|Vr_{our}-Vr_{Nord}|>5$~km~s$^{-1}$
we have a systematic difference 700~m~s$^{-1}$ and a r.m.s around the one-to-one relation
of 800~m~s$^{-1}$ for the FEROS radial velocities.
For the coud\'e measurements the mean difference is about 1.4~km~s$^{-1}$ and the r.m.s about 1~km~s$^{-1}$.
These results show the very good agreement of our radial velocities with those
obtained with CORAVEL by \citet{Nord04} and also that the FEROS and coud\'e measurements can be combined
since the systematic shift between the two instruments is within the errors.

As a byproduct of the radial velocity measurements, projected rotational velocities $V\sin(i)$
were computed for the FEROS spectra using the calibration of \cite{melo01}.
For fast rotators ($V\sin~i\ga30$~km~s$^{-1}$) the final rotational velocities were derived
as in \cite{melo03}.
In Figure~\ref{fig:compnord} (bottom) we compare our rotational velocities  with those from \cite{Nord04}.
Multiple stars are shown as encircled symbols.
Again, the one-to-one relation is shown as a continuous line whereas the dashed-dotted lines
mark the $V\sin(i)$ errors of 2~km~s$^{-1}$ for $V\sin~i\le20$~km~s$^{-1}$ and
$10\%$ of the observed $V\sin(i)$ for $V\sin~i>20$~km~s$^{-1}$.
A closer investigation of the 6 deviant points (those several km~s$^{-1}$ away from the
one-to-one relation) show that based on the high $\sigma(V_r)$ given by \cite{Nord04},
these objects are likely to be multiple stars seen in blend.

\subsection{Spectral type, Li and $H_\alpha$}

\begin{figure}

  \resizebox{\hsize}{!}{\includegraphics[angle=270, bb=50 90 550 750, clip]{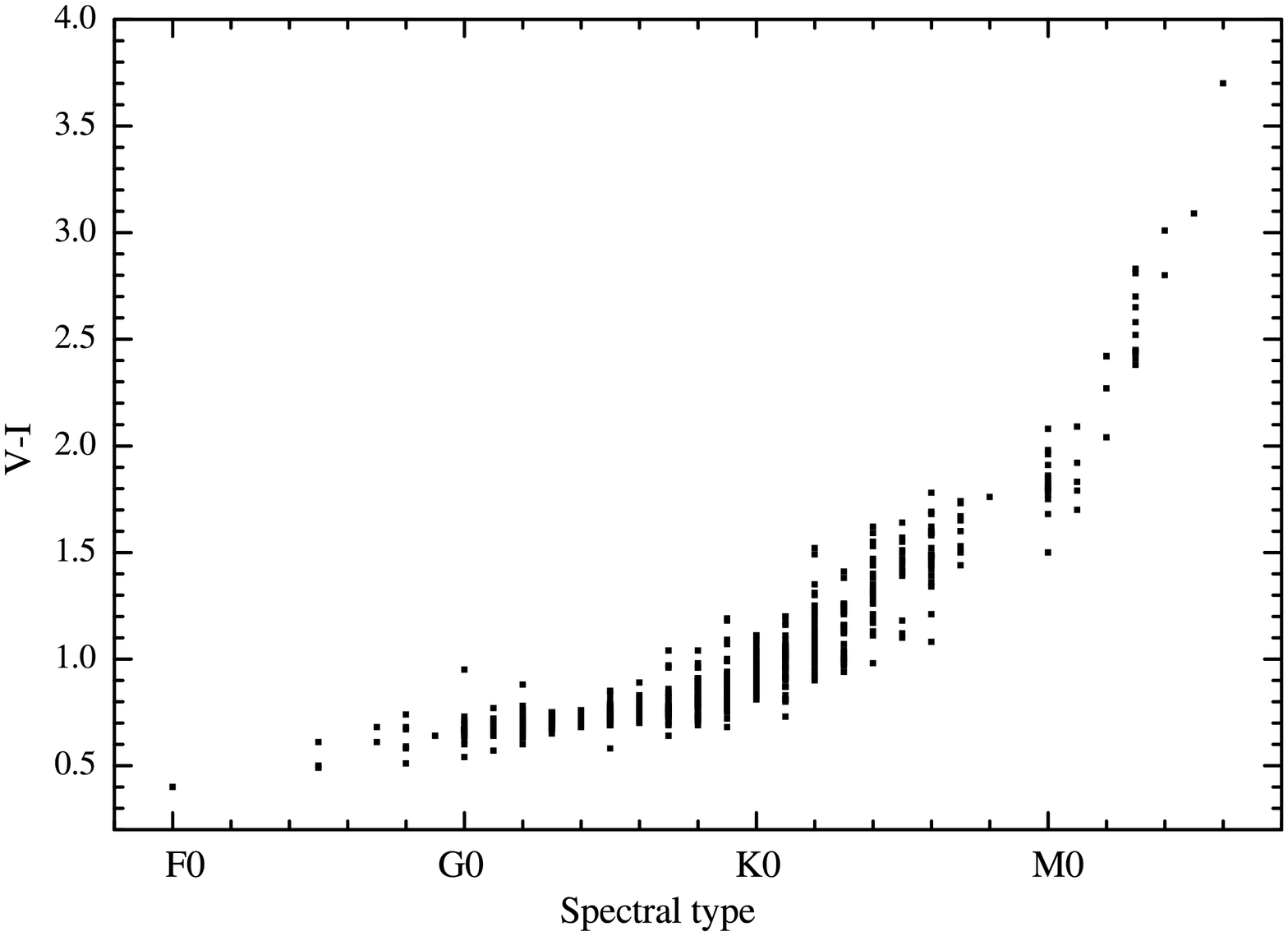}}
\caption{Our spectral types as a function of the $(V-I)_C$. }
\label{fig:st_color}
\end{figure}

\begin{figure}
  \resizebox{\hsize}{!}{\includegraphics[bb=50 269 530 775,clip]{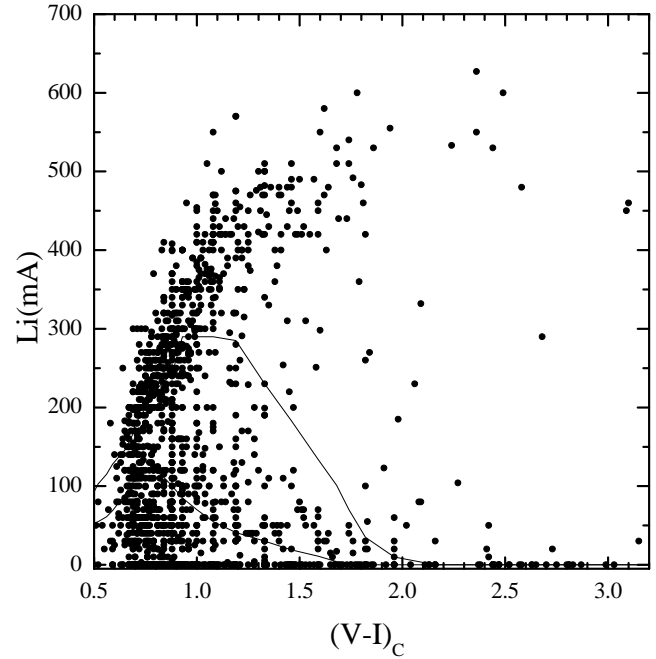}}
\caption{Distribution of Li equivalent width as a function of the $(V-I)_C$. 
The lines represents the upper and lower limits
for Pleiades members.}
\label{fig:li}
\end{figure}

Spectral classification was carried out for the observed spectra
using a combination of three different procedures:
i) comparison with the \cite{montes97} spectral library,
ii) calibrations of \cite{tripicchio97, tripicchio99} for the Na\,I~D and K\,I~7699~\AA\ lines and,
iii) the method developed by \cite{Torres1998} in the $H_\alpha$ region.
The good quality of the obtained  spectral types can be seen in Figure~\ref{fig:st_color}
where we plot the $(V-I)_C$ against our spectral types for dwarf stars.
We estimate the error in our spectral type classifcation to be about one sub-class.
Part of the spread in Figure~\ref{fig:st_color} is real, as we are dealing with young stars,
having distinct color excess, due to cool stellar spots, circumstellar material, etc.

Equivalent widths of the Li 6708~\AA\ and the $H_\alpha$ lines were measured from the 
FEROS or coud\'e spectra using IRAF {\it splot} task.
We estimate for low rotators ($V\sin~i\le20$~km~s$^{-1}$) observed with FEROS
an error of  $\la0.01$~\AA\ and twice for stars observed with the coud\'e.
The distribution of the Li equivalent width as a function of the $(V-I)_C$ is given in Fig.~\ref{fig:li}.
The large spread seen in Fig.~\ref{fig:li} is expected and it reflects the spread 
in age of the X-ray stars \citep{neu97}.
The combination of the Li equivalent widths and the color indexes was used to 
discriminate the sub-samples of {\it young} stars, the Pleiades age stars  and the old stars, 
as can be seen in the figure.

But in the search for young associations we will consider  "young star"
if the Li equivalent width is at least 90\% of that observed for
the Pleiades members  for a given $(V-I)_C$.
The few stars later than M0 with undetectable Li  were included in this young sub-sample
as the Li may be completely depleted for M stars already in the  PMS phase \citep{martin94}.

In summary,  from a sample of 1626 stars (1511 stars observed by us and 115 stars
taken from the literature), we classified 255
as giants stars and 1371
as dwarf  stars.
We considered as dwarfs also the subgiant stars (many of the young stars are
spectroscopically classified as subgiants), and 22 stars for which we did not obtain 
a luminosity class (mainly fast rotators or young stars taken from the literature).  
Among these 1371 dwarfs, 565, 282 and 524
are considered respectively younger, 
having similar age or older than the Pleiades.

\begin{figure*}
\begin{minipage}[c]{0.49\linewidth}
{\raggedleft\resizebox{\linewidth}{!}{
  \includegraphics[bb=108 575 495 789,clip]{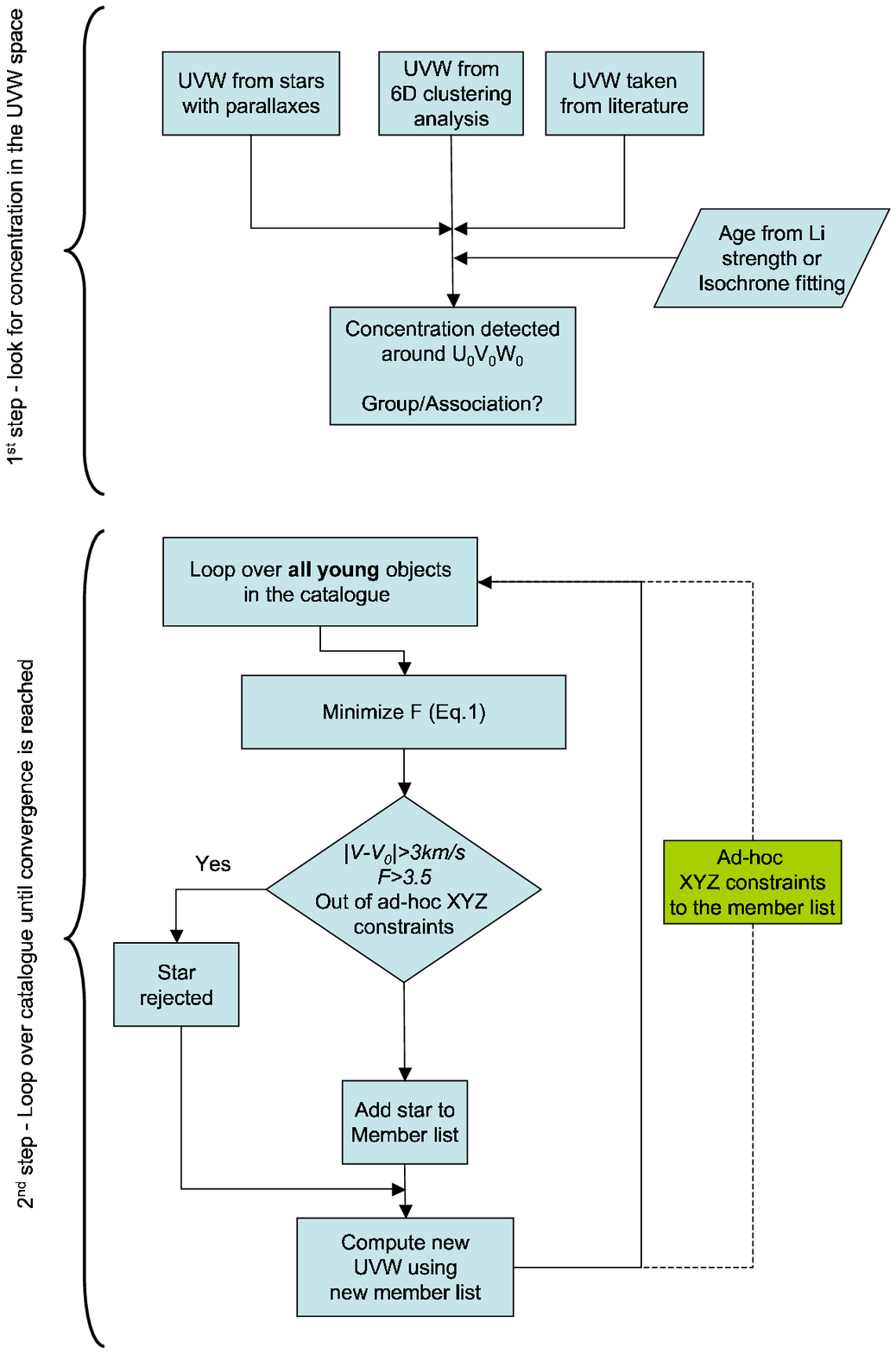}
 }}
\end{minipage}
\begin{minipage}[c]{0.49\linewidth}
{\raggedleft\resizebox{\linewidth}{!}{
\includegraphics[bb=108 222 474 574,clip]{5602fig5.eps} }}
\end{minipage}
\caption{The convergence method. {\it Left.} As a first step, a concentration around a point in the UVW--space is found.
The position of the stars defining this concentration at the HR diagram and their Li equivalent width give a first estimation
for the age of the association.
{\it Right.}
Starting with an age and an initial (UVW)$_0$ determined in step 1, we look for the stars in the young sample which fit the association.
}
\label{fig:flowchart}
\end{figure*}

\section{Finding associations: the kinematical method}

Intuitively one can consider an association as a group of stars appearing {\it concentrated}
together in a small area in the sky sharing some common properties such as age,
chemical composition, distance and kinematics. However, if such a group is close enough
to the Sun, its members will appear covering a large extent  in the sky
(as an example, Orion at 50~pc would cover almost the whole sky).
Thus, to find a group, projected spatial concentration
(i.e., in terms of right ascension and declination only)
and proper motions may not be enough.
A better criterion is to look for objects sharing
similar space motions (UVW) all around the sky (U positive towards the Galactic
center, V positive in the direction of Galactic rotation).
To obtain space velocities, proper motions ($\mu_{\alpha}$, $\mu_{\delta}$),
radial velocities and parallaxes are needed.
Since most of the  young stars observed in the SACY  have no parallaxes,
a convergence method to dynamically estimate their distances has been developed.
This method is described below and applied to the BPA in the next section.

\subsection{The kinematical convergence method}

As an initial step before applying the convergence method, we search for
an initial point (UVW)$_0$ in the UVW--space
around which a concentration of objects is observed indicating the existence of a possible association.
This can be done basically in three different ways:

i) considering only stars with known parallaxes;
ii) considering all stars with and without measured parallaxes;
iii) using reliable information from the literature.

i) In the first case, visual inspection of the distribution of objects in the UVW--space
can already reveal clustering around some points of the UVW--space.
These points may furnish a starting point, (UVW)$_0$, for the convergence method.

ii) In the second case, all stars are considered regardless whether they have parallaxes or not.
For those without parallaxes a distance is found assuming a common age for all objects.
(UVW) velocity components are then computed for all stars and as above, any
observed concentrations in UVW--space
can be  used as starting point.

iii) A third source for the (UVW)$_0$ may be the literature.
For instance, the starting  point for the BPA, presented in Section 6, can be found
either applying (i) or taking the (UVW) given by \cite{zuckermanSong04}.

An additional ingredient needed to start the convergence method is the age which is
estimated by the Li 6708~\AA\ strength and the isochrone fitting.
Once we have an (UVW)$_0$ and an age assigned we loop over all young stars of our sample.
For each star we search the distance $d$ which minimizes the quantity $F(m_v,\mu_\alpha,\mu_\delta,V_r)$.
The function $F$  is a quadratic combination of the
{\it evolutionary} distance (i.e., distance  in magnitudes of a given star to the adopted isochrone)
and the {\it kinematic} distance which is the 
Cartesian distance in $km~s^{-1}$ of the (UVW) of a star to the starting point (UVW)$_0$:

\begin{eqnarray}
\label{eq:f}
\nonumber \lefteqn{F(m_v,\mu_\alpha,\mu_\delta,V_r;d)=}\\
&[p\times(M_v-M_{v,iso})^2+...\\
\nonumber &...(U-U_0)^2+(V-V_0)^2+(W-W_0)^2]^{1/2}
\end{eqnarray}
where $p$ is a constant weighting the importance of the evolutionary
distance with respect to the kinematic distance. The kinematic distance is, in other words, 
the modulus of the velocity vector
\begin{equation}
\label{eq:vec}
d_{k}=|\nu-\nu_0|=[(U-U_0)^2+(V-V_0)^2+(W-W_0)^2]^{1/2}
\end{equation}

A limit of 3~km~s$^{-1}$ (i.e., $|\nu-\nu_0|<3~km~s^{-1}$) corresponding to 3 times 
the typical velocity dispersion of open clusters is set as kinematical criterion along 
with  $F~<~3.5$ which corresponds approximately to an evolutionary distance of less than 2
mag. If both conditions are observed the star is classified as possible member.

The procedure is then repeated until the convergence in the mean (UVW) is reached and the member list does not change any further.

As the main concentrations in UVW--space
are also spatially localized, albeit some of them covering large areas in the sky,
the distribution of the objects in physical space (XYZ) is used as an additional constraint.
In some cases similar concentrations in (UVW) split
up completely in the XYZ-space \citep{torres03}.

In general, {\it ad-hoc} limits in (XYZ) have to be imposed to avoid spurious stars and, for
nearby associations in UVWXYZ--space, mutual contaminations.
In these cases, the procedure is repeated with the new spatial constraints.
A summary of the description given above is schematically presented
as a flow-chart in Figure~\ref{fig:flowchart}.

It is worth noticing that due to the way our starting sample was defined (c.f., Sec. 2)
the members of the associations found by the SACY are a sub-sample
(or a lower limit) of the full list of members.

\subsection{Membership probabilities}

We  apply an alternative, objective, method to assess
membership probabilities of the stars in associations suggested by the
kinematical convergence analysis. We employ a {\sl predictive
classification} scheme that assumes group membership a priori for
a certain sub-sample of objects, the so-called training set. 
For example,  we can assume a training set derived by the kinematical
convergence method.

Our method works as follows: For each object, we standardize all 6
variables UVWXYZ by subtracting a mean value and dividing by a
standard deviation in order to guarantee similar weighting of each
component in the 6-dimensional space. Means and standard
deviations are derived from the training set only.  Next, we
apply a distribution free {\sl k-NN} (k nearest neighbors)
analysis to the entire sample to assign probabilities for each
object. This approach dispenses with the need to introduce a
probability density function. $k-NN$ probabilities are calculated
by defining euclidean spheres around each object in the 6
dimensional space that contains exactly $k$ objects. These $k$
objects contain $n_t$ objects from the training set, and
(trivially) $k-n_t$ other objects. The membership probability
$\alpha$ is then simply defined as $\alpha=n_t/k$. $\alpha$ is
defined for each point in the 6-dimensional UVWXYZ--space, and
changes discontinuously in steps of $1/k$. This method is, for
example, used to analyze the large-scale spatial distribution of
X-ray selected TTS from the ROSAT survey \citep{Sterzik95}, and
more details can be found in \cite{murtagh87}.

The use of the convergence method to define a {\it bona-fide} list of members is
independent of the assessment of  membership probability described here. Therefore, potential members
suggested by other works whose radial velocity and/or parallaxes have been collected from the literature
might be tested in similar way to those selected by the convergence method.

\section{Results: the $\beta$~Pic Association}

As a test case for the convergence method described in the previous section, we look for
the known BPA. As a starting point,
only young stars with trigonometric parallaxes are taken into account. Their UVW velocities show 
a concentration at U=-10~km~s$^{-1}$, V=-16~km~s$^{-1}$, W=-9~km~s$^{-1}$ which is
very close to the space motions found for the
BPA defined by \cite{zuckermanSong04}.

A few proposed BPA members in \cite{zuckermanSong04} not a priori in the SACY sample (mainly A-F stars 
excluded predominantly due to the temperature criterion)
were also included in our present analysis in order to test the consistence of the association as a whole, i.e.,
early- and late-type members. In addition, membership of 
a few {\it ad-hoc} stars not in \cite{zuckermanSong04} or in the SACY sample were
also tested, for example, the case of the isolated cTTS V4046 Sgr.
For all faint members of visual binaries, even wide, with no Hipparcos or Tycho-2 data we used
the proper motions and/or parallaxes of the primary star.

In the convergence method we do not use the Hipparcos parallaxes when the errors are
larger than 2~mas, as this may indicate unreliable parallaxes.
Anyway, these stars coud be used to check the resulting kinematical parallax.

\subsection{Observational isochrone}

\begin{figure}[bh]
  \resizebox{\hsize}{!}{\includegraphics{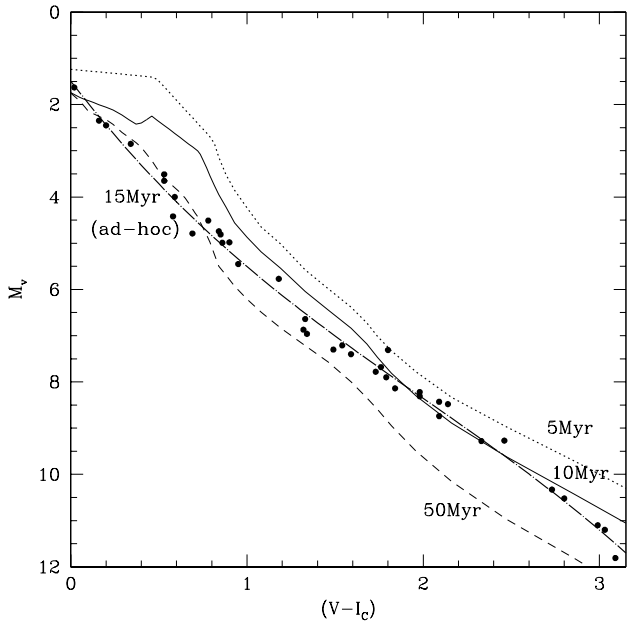}}
\caption{HR diagram from the proposed members of the $\beta$~Pic Association.
The isochrones over-plotted are those from \cite{siess00}.
The dotted and dashed lines correspond to the isochrones for 5~Myr and 50~Myr, respectively.
The isochrone for 10~Myr is shown as thick solid line whereas the
 observational isochrone is represented as a thick dotted-dashed line.}
\label{fig:hr_bpic}
\end{figure}

According to \cite{zuckerman01}, the BPA is 12$^{+8}_{-4}$~Myr old, and \cite{feige}
estimate the age of the BPA member GJ~3305 to be  13$^{+4}_{-3}$~Myr.
This is consistent with the 11~Myr  obtained by the dynamical tracing back models
\citep{ortega02, ortega04}.

The color-magnitude diagram with the isochrone of 10~Myr of \cite{siess00} is shown
in Figure~\ref{fig:hr_bpic}.
In contrast to low mass stars which agree  with the position of the isochrone,
stars earlier than about G5 are up to 2 magnitudes off from the theoretical predictions.
This discrepancy was already noticed by \cite{zuckerman01} who overcame the problem by
adopting observational isochrones based on stellar populations with well defined ages.
If on one hand isochrones are an important element to the convergence method applied here,
entering in the estimation of the evolutionary distance (see equation~\ref{eq:f}), on the other hand,
the temperature region where the differences between theory and observations seems more severe,
is mainly beyond the temperature cut-off applied to the SACY sample.

In spite of that, in order to be able to use the convergence method in future works covering a
broader temperature range, we adopted a similar strategy as \cite{zuckerman01}
and use the BPA members of Table~\ref{table:betapic}  to define an {\it ad-hoc} isochrone  of around 15Myr.
Our best fit gives (shown as a thick dot-dashed line in Figure~\ref{fig:hr_bpic}):

\begin{equation}
\label{eq:iso}
M_v = 1.50 + 4.98 (V-I)_C - 1.16 (V-I)_C^2 +0.193(V-I)_C^3
\end{equation}
for $-0.1 < (V-I)_C < 3.1$

Eq.~\ref{eq:iso} is then the isochrone used in the convergence method. At a first glimpse, the way this observational
isochrone was defined seems to lead to a circular argument, namely, the use Eq.~\ref{eq:iso} into the convergence method to
find the members which are going in turn to be used to refine Eq.~\ref{eq:iso}. It is worth noticing that the objects used to
set the early-type side of the observational isochrone all have determined parallaxes and are anchored in the UVW space and on
the color-magnitude diagram. The intrinsic assumption is that these early-type objects are indeed members of BPA. According to
our membership probabilities given in Table~\ref{table:betapic}, all but one A and F stars taken from \cite{zuckermanSong04}
have a membership probabilities higher than 85\% and are considered to be {\it bona-fide} members which validates our approach to
define the observational isochrone.

\subsection{Evidence for expansion}

\begin{figure}
  \resizebox{\hsize}{!}{\includegraphics{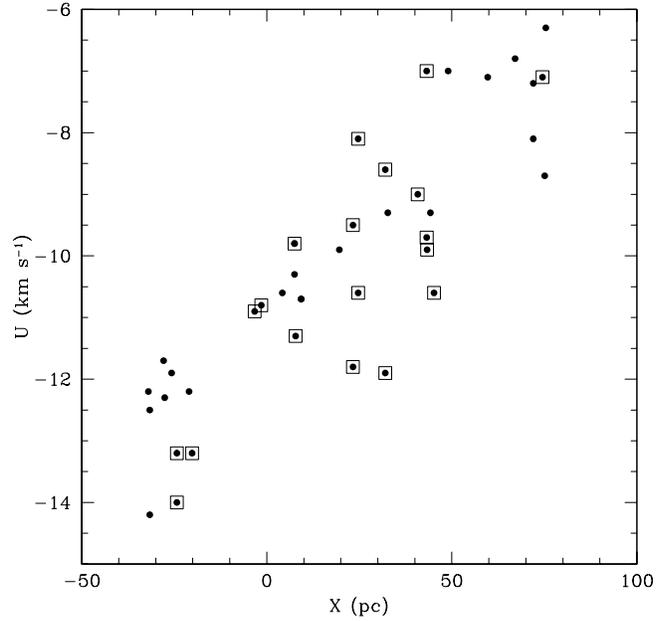}}
\caption{Correlation between the X-position and U-component of space velocity vector showing an expansion of the
members of the $\beta$~Pic Association .
Objects whose parallaxes are from Hipparcos Catalogue are overplotted with squares.}
\label{fig:expansion}
\end{figure}

\begin{figure*}
  \begin{center}
  \begin{tabular}{c}
   \resizebox{0.98\hsize}{!}{\includegraphics{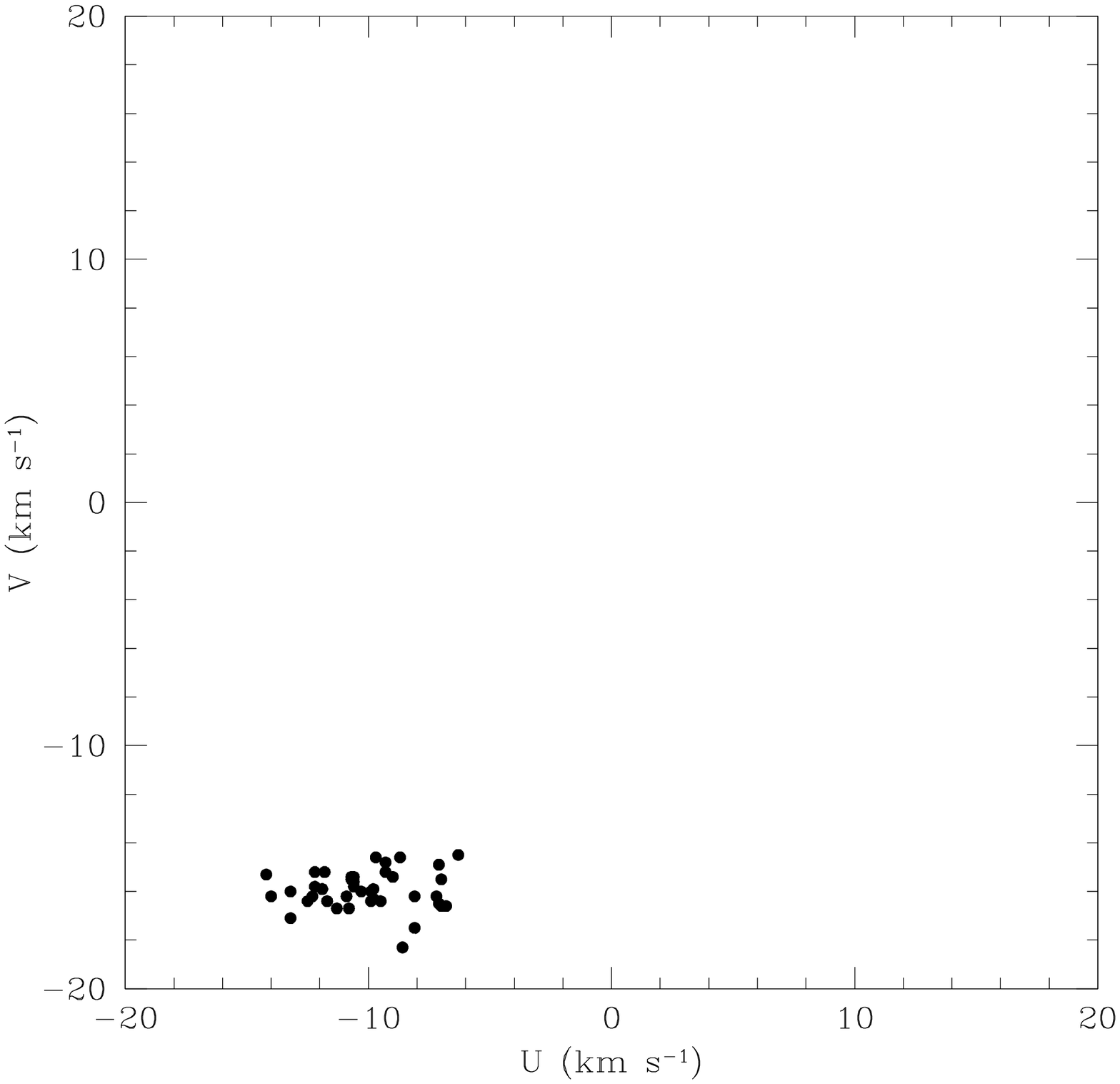}\includegraphics{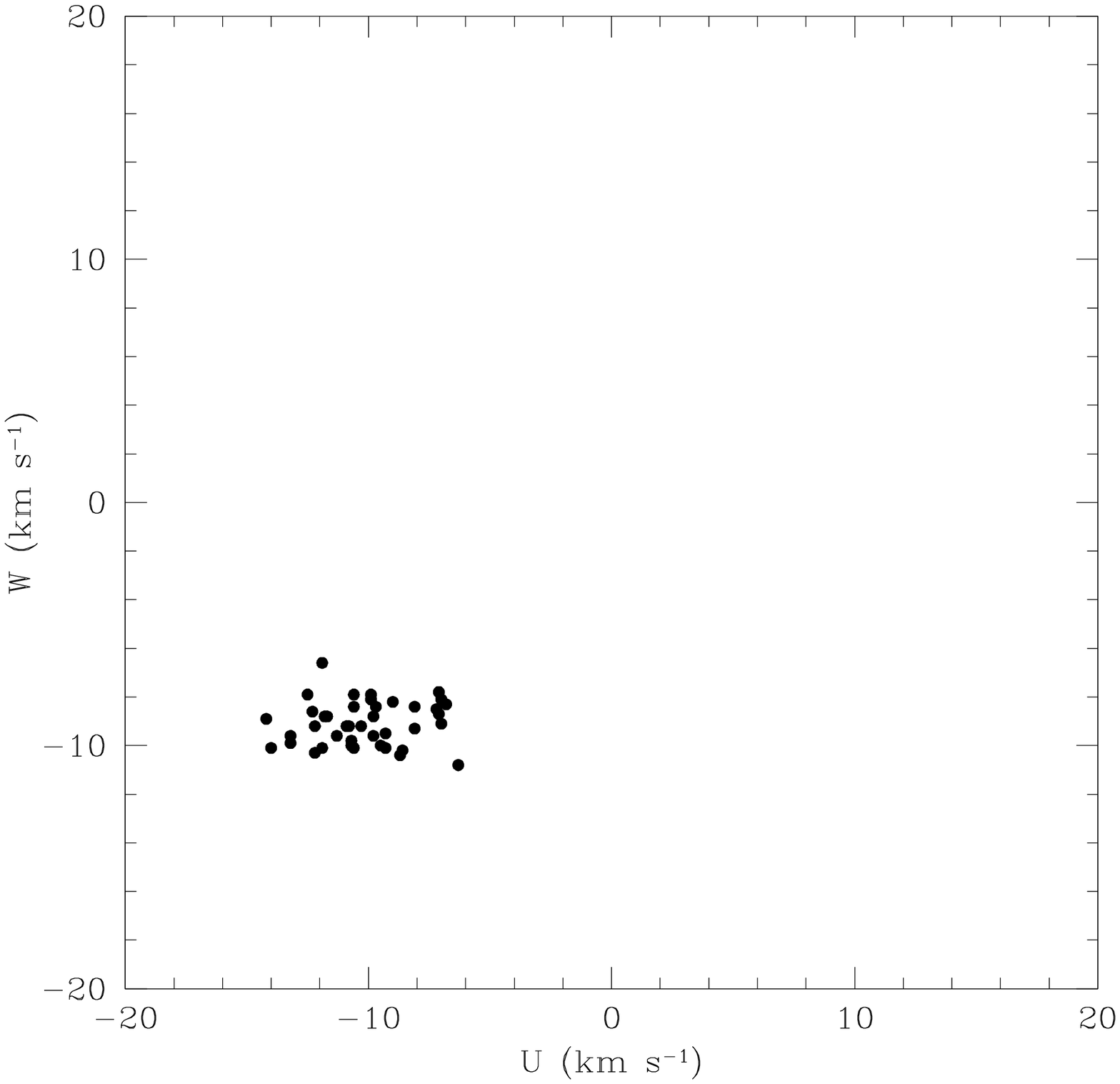}\includegraphics{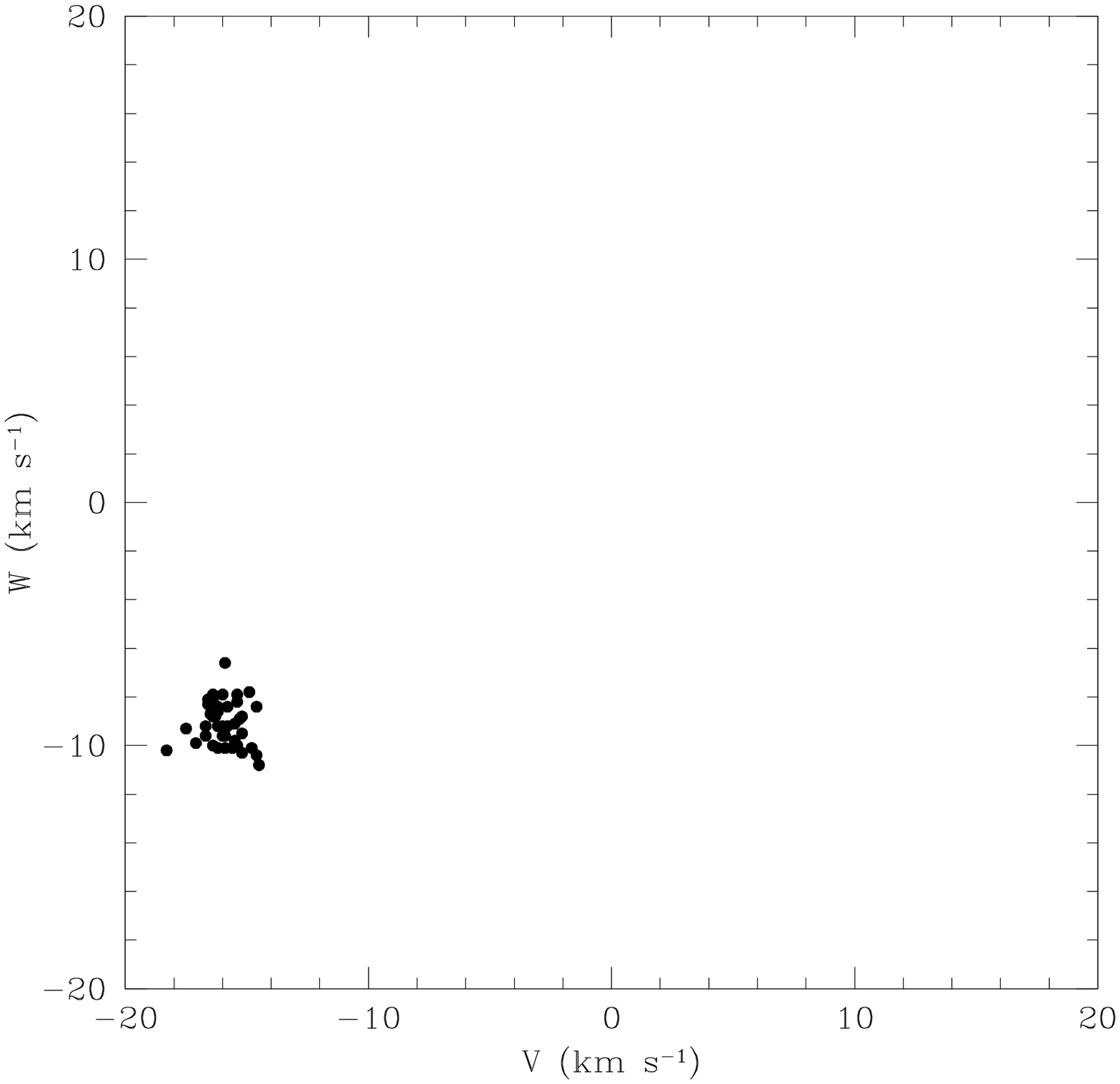}}\\
   \resizebox{0.98\hsize}{!}{\includegraphics{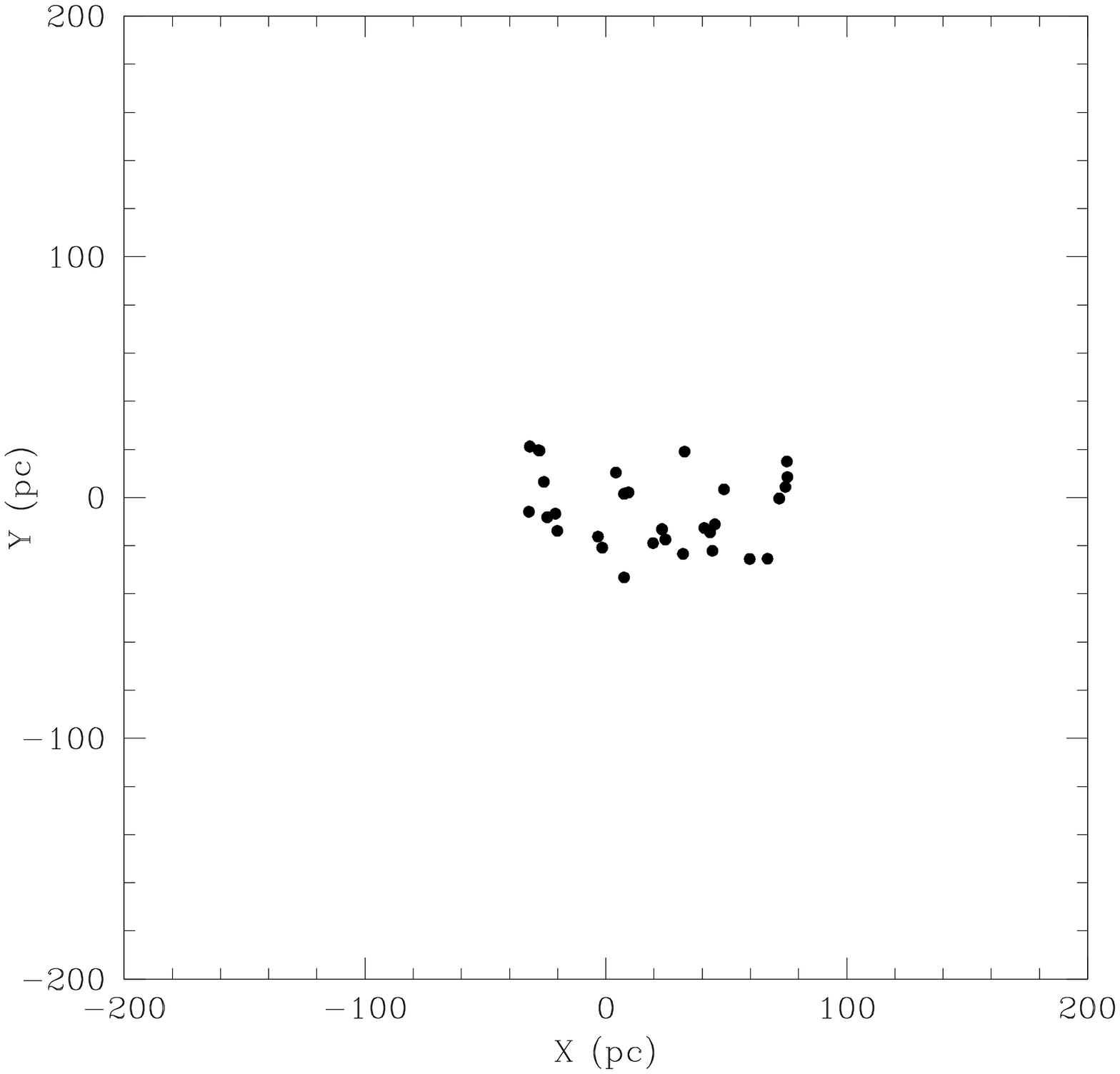}\includegraphics{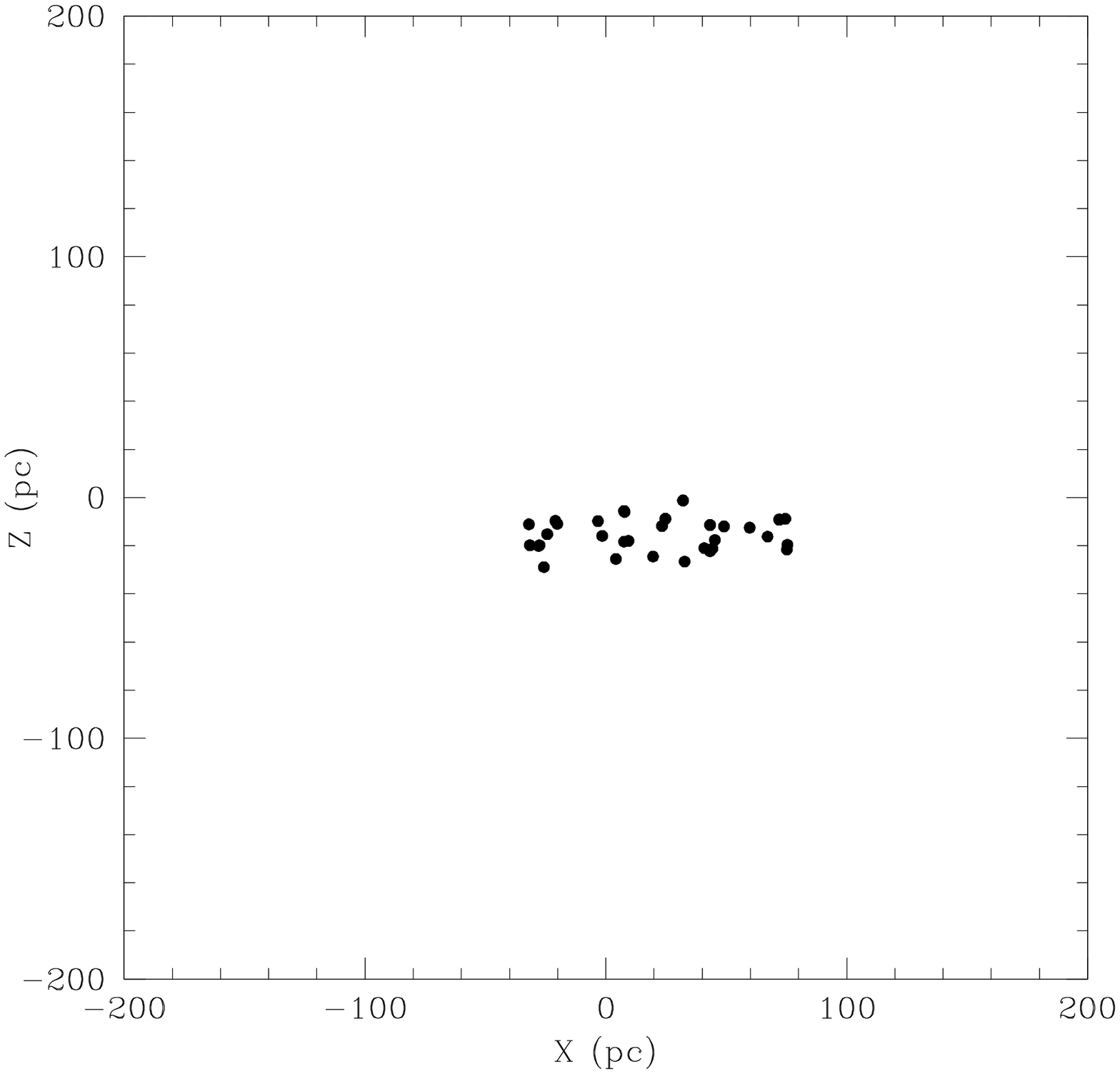}\includegraphics{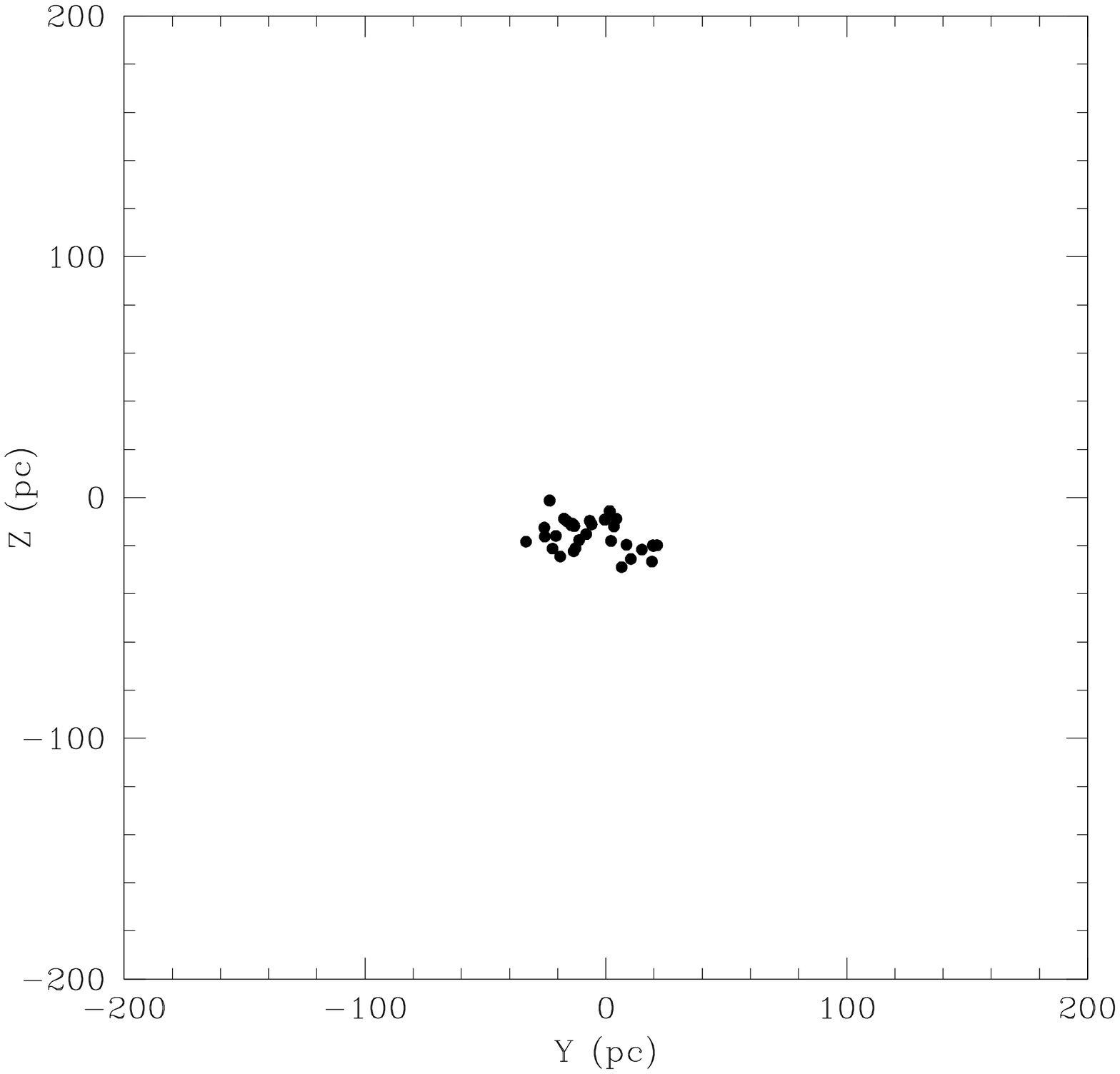}}\\
  \end{tabular}
    \end{center}
\caption{Combinations of the sub-spaces of the UVWXYZ--space showing
a well defined clustering in both kinematical and spatial coordinates.}
\label{fig:xyz_uvw}
\end{figure*}

A plot of the velocity component U of the space velocity vector against 
the space direction X is seen in Figure~\ref{fig:expansion}. 
The farthest components show a larger U-component velocity suggesting that the association is actually expanding. 
A quick check of whether this effect is an artifact of our convergence method can be carried out looking 
for the objects whose parallaxes were taken from Hipparcos and {\it not} derived by  the convergence method. 
These objects are overplotted with squares in Figure~\ref{fig:expansion}. 
It is clear that the correlation does exist, even if the members are restricted to 50~pc as is the
case of the \cite{zuckermanSong04} analysis. A linear least-squares fit gives:

\begin{equation}
\label{eq:exp}
U = 0.053 (X) -11.1
\end{equation}

Such a correlation seems to be a common trend observed among the young stars (i.e., $age\la age_{Pleiades}$) 
of the SACY sample in contrast to the older sample where no correlation is seen. 
In addition, no  significant correlation seems to exist either between Y and V or W and Z, 
regardless of the age of the objects. 
Interpretation of what caused the trend seen in Figure~\ref{fig:expansion} and
its possible links to star formation and early dynamic interaction is postponed for a future paper.

In the convergence method described in Sec. 5.1, Eq.~\ref{eq:exp} had to be taken 
into account regardless the physical origin of this expansion.
This is done simply by replacing the $U_0$ in Eq.~\ref{eq:f} by an $U_0^\prime$ such that
$U_0^\prime = U_0 + kX$.

\subsection{The $\beta$~Pic Association membership probabilities}

We estimated the membership probabilities
of all possible members to the BPA  as discussed in Sec. 5.2. We assume the training set as derived 
by the kinematical convergence method, given in Table~\ref{table:betapic}.

For the BPA analysis we assume $k=20$. {\it All}
members of the training set have probabilities in the range of
0.7-1.0, i.e. they are already well confined and clustered in the
UVWXYZ--space. Thus,{\it bona-fide} members in this list have
membership probability of, at least, 0.75.  
These probabilities are displayed in the last column
in Table~\ref{table:betapic}.

We do find four stars above this probability threshold which are not
given in Table~\ref{table:betapic}.  
Three of them have magnitude
differences with respect to the observational isochrone (Eq.~\ref{eq:iso})
larger than 1.4~mag and therefore were discarded. 
The other is Hip~88726, having (X, U) values 
(42.3pc, -13.0 km s$^{-1}$),
not compatible with the observed trend seen in Fig.~\ref{fig:expansion} (or Eq.~\ref{eq:exp}).
We rejected it as a BPA member in spite of a membership 
probability of 90\% and an acceptable magnitude.
\cite{song03} also considered it as an outlier.

The final list of  high probability proposed members is  the same
as that obtained by the convergence method.

\subsection{The SACY $\beta$~Pic Association}

The final results for the BPA are summarized  in Table~\ref{table:betapic},
where the magnitudes and colors are corrected by the presence of companions
as explained in Section~2.2.

Table~\ref{table:betapic} contains 41 highly probable members, that could be used to define
the main properties of the BPA and 2 possible members.

\cite{zuckermanSong04} proposed a list of 33 members\footnote {Actually their list has 34 entries, 
but 2 entries are for both members of the SB2 V824~Ara.} to the BPA including 
stars hotter than the SACY limit of $B-V\geq0.6$ and
in both hemispheres.
Only one very faint star proposed as member by Zuckerman \& Song was
not used in our analysis as it lacks the kinematical data needed to be injected in the convergence method.

\setcounter{table}{5}

\begin{landscape}
\begin{table}
 
\caption{Data of the proposed members of the $\beta$ Pic Association. Column 1: Star Identification; Column 2 and 3: Equatorial Coordinates; Column 4 and 5: Proper Motions;
Column 6: Radial Velocities and Notes about duplicity; Column 7: Rotational Velocities; Column 8 and 9: Photometry; Column 10: Spectral Types; Column 11: Li Equivalent Widths; Column 12: Trigonometric Parallaxes; Column 13: Kinematical Parallaxes; 
Column 14 to 16: Space Velocities; Column 17 to 19: Space Positions; Column 20: Function F (see text); Column 21: Membership Probabilities; Column 22: Notes. }
{\tiny
\label{table:betapic}
\begin{tabular}{lccrrlrlrlrrrr@{\hspace{1.5pt}}r@{\hspace{1.5pt}}rr@{\hspace{1.5pt}}r@{\hspace{1.5pt}}rccl}
\hline

Name  &       $\alpha$&$\delta$&$\mu_\alpha$&$\mu_\delta$&$~~~V_r$&Vsini&$V_J$ & V-I$_C$ &SpT &E$_{Li}$& $\pi_{tri}~~~$    & $\pi_{kin}$&\multicolumn{3}{c}{$(U,V,W)$}&\multicolumn{3}{c}{$(X,Y,Z)$}&F&Prob&notes\\
     &          2000&2000 &\multicolumn{2}{c}{~~mas yr$^{-1}$}&\multicolumn{2}{c}{~km s$^{-1}$}&&	  &    &m\AA &mas~~~	    &  mas     &\multicolumn{3}{c}{km s$^{-1}$}&\multicolumn{3}{c}{pc}     & &    &	 \\
\hline\hline
Hip 10679	&02 17 24.7	&+28 44 31	&86.7	&-76.7	&~{\it5.0$\pm$0.4}	&{\it 8~~}    &~{\it 7.75}	&{\it 0.69}	&{\it G2V}   &{\it 160}	&29.4$\pm5.4$   &25.5	&-12.3 & -16.2 & -8.6	&-27.6 & 19.5 & -19.8	&1.03	&1.00 &Z,s,N,H,3\\
HD 14082	&02 17 25.2	&+28 44 43	&84.3	&-77.6	&~{\it4.6$\pm$0.3}	&{\it 45~~}   &~{\it 6.99}	&{\it 0.59}	&{\it F5V}   &{\it 140}	&25.4$\pm$2.8   &25.3   &-11.7 & -16.4 & -8.8	&-27.9 & 19.7 & -20.1	&1.05	&1.00 &Z,N,H,3\\
BD+30 397B	&02 27 28.1	&+30 58 41	&79.5	&-70.1	&~{\it4.7$\pm$1.3}	&             &{\it 12.44}	&{\it 2.33}	&{\it M2Ve}  &{\it 110}	&               &23.3	&-12.5 & -16.4 & -7.9	&-31.6 & 21.2 & -19.8	&1.39	&1.00 &Z,s,3,5,6\\
AG Tri  	&02 27 29.3	&+30 58 25	&79.5	&-70.1	&~{\it7.0$\pm$1.1}	&{\it 5~~}    &{\it 10.12}	&{\it 1.34}	&{\it K6Ve}  &{\it 220} &23.7$\pm$2.0   &23.3	&-14.2 & -15.3 & -8.9	&-31.6 & 21.2 & -19.8	&1.81	&1.00 &Z,s,H,3,4\\
BD+05 378	&02 41 25.8	&+05 59 19	&82.3	&-55.1	&{\it10.0} SB1	        &             &{\it 10.37}	&{\it 1.59}	&K6Ve	     &450       &24.7$\pm$2.4   &25.5	&-11.9 & -15.9 & -6.6	&-25.7 & 6.5  & -28.9	&2.62	&0.85 &Z,S,3,7\\
HD 29391	&04 37 36.1	&-02 28 25	&43.3	&-64.2	&{\it21.0}   	        &{\it 95~~}   &~{\it 5.22}	&{\it 0.34}	&{\it F0V}   &{\it 0}	&33.6$\pm$0.9	&       &-14.0 & -16.2 & -10.1  &-24.3 & -8.2 & -15.2	&1.95	&0.90&Z,2,1\\
GJ 3305	        &04 37 37.5	&-02 29 28	&46.1	&-64.8	&{\it20.1}   	        &             &{\it 10.59}	&{\it 1.98}	&{\it M1Ve}  &{\it 120}	&33.6$\pm$0.9   &    	&-13.2 & -16.0 & -9.6	&-24.3 & -8.2 & -15.2	&1.01	&0.90&Z,9,10,11\\
V1005 Ori	&04 59 34.8	&+01 47 02	&37.2	&-93.9	&{\it18.7}		&{\it 14~~}   &{\it 10.05}	&{\it 1.84}	&M0Ve        &270       &37.5$\pm$2.6   &41.5	&-12.2 & -15.8 & -9.2	&-21.0 & -6.7 & -9.7	&0.53	&0.90&S,H,10,13\\
CD-57 1054	&05 00 47.1	&-57 15 26	&35.6	&72.8	&19.4$\pm$0.3	        &5.4          &10.00	        &1.79	        &M0Ve	     &360       &38.1$\pm$1.1	&	&-10.8 & -16.7 & -9.2	&-1.5  &-20.8 & -15.9	&0.91	&0.90&Z,S\\
Hip 23418       &05 01 58.8     &+09 59 00      &17.2   &-82.0  &{\it17.3} SB2          &             &{\it 11.95}*     &{\it 2.46}     &{\it M3Ve}  &{\it 0}   &31.2$\pm$8.6   &29.1   &-12.2 &-15.2  &-10.3   &-32.0 &-5.9  &-11.1    &1.77   &0.90&Z,s,3,22,8\\     
HD 35850	&05 27 04.7	&-11 54 03	&17.2	&-49.3	&{\it22.8} SB2	        &{\it 50~~}   &~{\it 6.56}*	&{\it 0.58}	&{\it F7V}   &{\it 191}	&37.3$\pm$0.8   &	&-13.2 & -17.1 & -9.9	&-20.2 &-13.8 & -10.9	&1.97	&0.90&Z,N,H,14\\
$\beta$ Pic	&05 47 17.1	&-51 04 00	&4.7	&82.0	&{\it20.2$\pm$0.4}	&{\it 139~~}  &~3.77	        &0.16	        &{\it A3V}   &{\it 0}	&51.9$\pm$0.5	&	&-10.9 & -16.2 & -9.2	&-3.3  & -16.2& -9.8	&0.55	&0.90&Z,1\\
AO Men	        &06 18 28.2	&-72 02 42	&-8.1	&71.4	&16.3   	        &16.4         &~{\it 9.80}	&{\it 1.32}	&K4Ve	     &420	&25.9$\pm$0.9	&	&-9.8  & -16.3 & -8.8	&7.5   & -33.2& -18.3	&1.45	&0.90&Z,S,21\\
HD139084B	&15 38 56.8	&-57 42 19	&-52.9	&-106.0	&~0.1$\pm$2.0    	&             &~{\it 14.80}     &{\it2.90}      &M5Ve        &460       &25.2$\pm$1.1   &	&-11.9 &-15.9  & -10.1  &32.0  & -23.4& -1.3	&2.71	&0.85&Z,S,3\\
V343 Nor 	&15 38 57.6	&-57 42 26	&-52.9	&-106.0	&~4.2$\pm$1.4    	&16.6         &~7.97	        &0.90	        &K0V	     &292       &25.2$\pm$1.1   &	&-8.6  &-18.3  & -10.2  &32.0  & -23.4& -1.3	&2.85	&0.80&Z,S\\
V824 Ara	&17 17 25.5	&-66 57 02	&-21.8	&-136.5	&~{\it5.9} SB2	        &31~~         &~{\it 7.23}*	&{\it 0.84}	&G7IV        &250	&31.8$\pm$0.7	&	&-8.1 & -17.5 & -9.3	&24.7  & -17.4& -8.8	&2.43	&0.85&Z,S,15,19\\
HD155555C       &17 17 31.3     &-66 57 06      &-21.8  &-136.5 &~2.7$\pm1.8$           &6~~          &{\it12.82}       &{\it 2.73}     &M3Ve        &20:       &31.8$\pm$0.7   &       &-10.6 &-15.8 & -8.4    &24.7  & -17.4& -8.8    &1.11   &0.90&Z,S,20\\
CD-54 7336	&17 29 55.1	&-54 15 49	&-7.0	&-63.1	&~1.6$\pm$1.4	        &35.3         &~9.55	        &0.95	        &K1V	     &360       &               &15.1	&-7.1  & -16.5 & -8.7	&59.7  & -25.5& -12.5	&1.14	&0.85&S\\
HD 161460	&17 48 33.8	&-53 06 43	&-3.9	&-56.4	&~1.4 SB2	        &13.6         &~9.32*	        &0.86	        &K0IV        &320       &               &13.6	&-6.8 & -16.6  & -8.3	&67.1 & -25.4 & -16.2	&1.34	&0.80&S\\
HD 164249	&18 03 03.4	&-51 38 56	&3.5	&-86.5	&~0.5$\pm$0.4	        &{\it 20~~}   &~{\it 7.01}      &{\it 0.53}	&F6V         &107     	&21.3$\pm$0.9   &	&-7.0 & -15.5 & -9.1	&43.2 & -14.4 & -11.4	&1.95	&0.95&Z,H,1\\
HD 164249B	&18 03 04.1	&-51 38 56	&3.5	&-86.5	&-2.4$\pm$1.3	        &             &12.5:            &          	&M2Ve        &70      	&21.3$\pm$0.9   &	&-9.7 & -14.6 & -8.4	&43.2 & -14.4 & -11.4	&1.90	&0.95&\\
V4046 Sgr	&18 14 10.5	&-32 47 33	&2.1	&-54.5	&-6.9 SB2	        &13.8         &10.94*	        &1.33	        &K6Ve	     &440       &               &13.8	&-8.1 & -16.2 & -8.4	&72.0 & -0.4 & -9.1	&1.13	&0.90&16\\
G7396-0759      &18 14 22.1	&-32 46 10	&2.1	&-54.5	&-5.7 SB?               &3.0          &12.78	        &2.14	        &M1Ve	     &200       &               &13.8	&-6.9 & -16.2 & -8.6	&72.0 & -0.4 & -9.2	&0.99	&0.85&\\
HD 168210	&18 19 52.2	&-29 16 32	&2.8	&-47.2	&-7.0$\pm$2.6	        &102.7        &~8.89	        &0.78	        &G5V	     &290	&13.3$\pm$1.4	&	&-7.1 & -14.9 & -7.8	&74.5 & 4.4  & -8.8	&1.78	&0.85&S\\
HD 172555	&18 45 26.9	&-64 52 15	&32.7	&-148.7	&~{\it3.8}		&{\it116}~~   &~{\it4.78}       &{\it0.20}      &{\it A6IV}  &{\it0}	&34.2$\pm$0.7   &	&-9.5 & -16.4 & -10.0	&23.3 & -13.2 & -11.8	&1.08	&0.85&Z,H,17\\
CD-64 1208      &18 45 36.9	&-64 51 48	&32.7	&-148.7	&~1.0$\pm$3.0           &{\it100:}~   &~{\it9.54}       &{\it1.54}      &K5Ve        &490   	&34.2$\pm$0.7   &	&-11.8& -15.2 & -8.8 	&23.3 & -13.1 & -11.8	&2.04	&0.95&Z,1\\
T9073-0762	&18 46 52.6	&-62 10 36	&18.1	&-76.6	&~2.4$\pm$0.1	        &9.9          &12.08	        &2.09	        &M1Ve	     &332	&               &18.6	&-9.3 & -15.2 & -9.5	&44.2 & -22.1 & -21.2	&1.01	&0.90&S\\
T7408-0054 	&18 50 44.5	&-31 47 47	&10.6	&-77.8	&-6.0$\pm$1.0           &49.7         &11.20	        &1.76	        &K7Ve	     &492       &               &19.8	&-7.0 & -16.6 & -8.1	&49.0 & 3.4   & -12.0	&1.95	&0.90&S\\
PZ Tel  	&18 53 05.9	&-50 10 49	&16.6	&-83.5	&-3.4$\pm$0.7	        &69.0         &~8.29	        &0.85	        &G9IV	     &287	&20.1$\pm$1.2	&	&-10.6 & -15.4 & -7.9	&45.2 & -11.1 & -17.6	&2.34	&0.95&Z,S\\
T6872-1011      &18 58 04.2	&-29 53 05      &16.1	&-47.2	&-4.9$\pm$1.0	        &33.8         &11.78	        &1.80	        &M0Ve	     &483	&            	&12.8   &-6.3 & -14.5 & -10.8   &75.4 &   8.5 & -19.6	&2.68	&0.75&S\\
CD-26 13904	&19 11 44.6	&-26 04 09	&18.9	&-44.2	&-8.1$\pm$0.3    	&9.8          &10.27*	        &1.18	        &K4V	     &320       &               &12.6	&-8.7 & -14.6 & -10.4	&75.1 & 15.0 & -21.6	&2.50	&0.85&S\\
$\eta$ Tel	&19 22 51.2	&-54 25 25	&25.6	&-83.0	 &~{\it0} ~~D4.1"	&{\it330:~}   &~5.02	        &-0.02   	&{\it A0V}   &{\it0}    &21.0$\pm$0.7   &	&-9.0 & -15.4 & -8.2	&40.8 & -12.6 & -21.0	&1.17	&0.95&Z,3,18\\
HD 181327	&19 22 58.9	&-54 32 16	&23.8	&-81.8	&-0.7           	&{\it16~~}    &~{\it7.03}       &{\it0.53}	&F6V         &120       &19.8$\pm$0.8   &	&-9.9 & -16.0 & -7.9	&43.3 & -13.4 & -22.3	&1.75	&0.90&Z,H,1\\
AT MicN  	&20 41 51.1	&-32 26 07	&269.3	&-365.7	&-4.5      	        &10.1         &10.99*	        &2.99	        &M4Ve	     &0	        &97.8$\pm$4.7   &105.1	&- 9.8 & -15.9 & -9.6	&7.5 & 1.5 & -5.6	&1.12	&0.90&Z,S\\
AT MicS  	&20 41 51.1	&-32 26 10	&269.3	&-365.7	&-5.2      	        &15.8         &11.09*	        &3.03	        &M4Ve	     &0	        &97.8$\pm$4.7   &105.1	&-10.3 & -16.0 & -9.2	&7.5 & 1.5 & -5.6	&0.56	&0.90&Z,S\\
AU Mic	        &20 45 09.5	&-31 20 27	&280.4	&-360.1	&-6.0$\pm$1.7           &9.3          &~8.73	        &2.09	        &M1Ve	     &80	&100.6$\pm$1.4  &	&-11.3 & -16.7 & -9.6 	&7.8 & 1.7 & -6.0	&1.15	&0.90&Z,S\\
AZ Cap  	&20 56 02.7	&-17 10 54	&59.3	&-63.0	&-6.9 D2.2"		&15.6         &10.62*	        &1.49	        &K6Ve	     &420       &               &21.6	&-9.3 & -14.8 & -10.1	&32.7 & 19.1 & -26.6	&1.65	&0.90&Z,S\\
CP-72 2713	&22 42 49.0	&-71 42 21	&94.1	&-54.4	&~8.6$\pm$0.5	        &7.5          &10.60	        &1.73	        &K7Ve	     &440       &               &27.3	&-9.9 & -16.4 & -8.1	&19.6 & -18.9 & -24.5	&1.18	&0.80&S\\
WW PsA	        &22 44 57.9	&-33 15 02	&183.1	&-118.9	&~2.2		        &12.1         &12.07 	        &2.80	        &M4Ve	     &0	        &42.4$\pm$3.4   &49.0	&-10.7 & -15.5 & -9.8	&9.3 & 2.1 & -18.0	&0.86	&0.95&Z,S\\
TX PsA 	        &22 45 00.0	&-33 15 26	&183.1	&-118.9	&~2.4		        &16.8         &13.36	        &3.09	        &M5Ve	     &450       &               &49.0	&-10.7 & -15.4 & -10.0	&9.3 & 2.1 & -18.0	&1.18	&0.90&Z,S\\
BD-13 6424	&23 32 30.9	&-12 15 52	&138.1	&-83.2	&~1.8$\pm$0.7           &8.8          &10.54	        &1.98	        &M0Ve	     &185	&               &35.7	&-10.6 & -15.7 & -10.1	&4.2 & 10.4 & -25.7	&1.00	&0.90&S\\
\hline
\multicolumn{22}{c} {\bf {possible members}}\\
\hline
\hline
HD 203	        &00 06 50.0	&-23 06 27	&97.3	&-47.1	&~9.7$\pm$2.0       	&{\it 155~~}  &~{\it 6.19}      &{\it 0.46}	&F3V	     &87	&25.6$\pm$0.8	&       &-10.4 & -14.5 & -13.3	&4.5   & 5.8  & -38.4	&4.53	&0.45 &Z,H,1\\
HD 199143	&20 55 47.7	&-17 06 51	&59.2	&-61.6	&-4.5$\pm$2.1   	&{\it155~~}   &~{\it7.35}*	&{\it0.60}	&F7V         &150       &21.0$\pm$1.0   &	&-7.8  & -13.9 & -11.8  &33.7 & 19.7 & -27.3	&3.72	&0.60&Z,H,12\\
\hline
\end{tabular}
}
\vfill
Data in italic were taken from the literature; 
(*)  photometric data for SB2 or close visual binaries are corrected taken into account the companions
(Hip~23418, CD-26~13904 and HD~199143 have  faint visual companions at $\sim1\arcsec$)
Z~=~proposed member by \citet{zuckermanSong04}; 
s~=~within SACY definition, {\it but in the Northern Hemisphere},  and not observed by us; 
S~=~observed in the SACY program;  H~=~photometry from Hipparcos; 
N~=~radial velocity and/or $V\sin~i$ from \citet{Nord04};  
(1) \citet{zuckerman01}; (2) \citet{barbier}; (3) \citet{song03}; (4) \citet{cutispoto00}; (5) \citet{weis91}; 
(6) \citet{ianna}; (7) \citet{robertson}; 
(8) \citet{koen}
(9) \citet{weis91b}; (10) \citet{gizis};
(11) \citet{feige}; 
(12) \citet{christian};
(13) \citet{favata95}; (14)  \citet{wichmann03}; 
(15) \citet{strass}; (16) \citet{quast00}; (17) \citet{erspamer}; (18) \citet{gray87};
(19)\citet{cutispoto98}; (20) \citep{eggen78}; (21) \citet{cutispoto99}; 
(22) \citet{delfosse}; (19)\citet{cutispoto98}; (20) \citep{eggen78}.
\end{table}
\end{landscape}

Four stars of their list are not proposed as probable members:
\begin{itemize}

\item {\it Hip~88726 and Hip~79881} - they do not fit into the expansion relation (Hip~79881
has a membership probability of 65\%)
and they were also rejected by \cite{song03}.

\item {\it HD~203 and HD~199143} - they are high rotational velocity F stars ($V\sin(i)$~=~155 km s$^{-1}$) and
their radial velocities are less reliable. Both stars would be considered as good members with radial velocity
values within one sigma of the mean used values or if we have used published radial velocities. 
We consider them as possible members.

\end{itemize}

From the 13 new members proposed in this paper only two have Hipparcos parallaxes
(HD~168210 and Hip~23200 = V1005~Ori). Another one is the new visual companion of HD~164249,
and we use the Hipparcos parallax of the primary.

In Figure~\ref{fig:xyz_uvw} we show  different projections in UVWXYZ--space.
In all panels a well defined clustering in both spatial and kinematical coordinates is seen which 
gives confidence of the solution obtained by the convergence method.
The clustering has an oblate shape along the X and U directions, as discussed in section 6.2.

According to Table~\ref{table:betapic}, the association
has a mean (UVW) of ($-10.1\pm2.1$~km~s$^{-1}$, $-15.9\pm0.8$~km~s$^{-1}$, $-9.1\pm0.9$~km~s$^{-1}$),
and a mean (XYZ) of ($20\pm34~$pc, $-5\pm15~$pc, $-15\pm7~$pc).
It spreads over a 100~pc interval in X, 50~pc in Y and 30~pc in Z. 
The distances of the proposed members of the BPA to the Sun range from 10~pc up to 80~pc.
If only stars later than G0 (28 stars) are considered
the mean (UVW) would barely change  ($-10.1\pm2.0$~km~s$^{-1}$, $-16.0\pm0.7$~km~s$^{-1}$ and $-9.1\pm1.0$~km~s$^{-1}$).
The distances would change less than 1\%.

It is evident that the solar position is almost inside the BPA
and thus it could have members spread  over all the sky, and
there may exist members undetected by our survey due to its partial sky coverage. 
Nevertheless, the distribution of the members in XZ-plane seems to indicate
that the  BPA is an association mainly in the Southern skies.
Even a very high negative X value, if we extrapolate the expansion using the mean Y and Z, 
would imply in stars near $\alpha~\sim~$05H and $\delta~\sim~+10\degr$.
To be sure where BPA ends at negative X direction it is essential to extend SACY at this region.
The gap near X~=~-10 in Figure~\ref{fig:expansion} comes from the incomplete sky coverage in this region.
Nevertheless, an extreme positive X value would fall within our sample.

\begin{figure}
  \resizebox{\hsize}{!}{\includegraphics{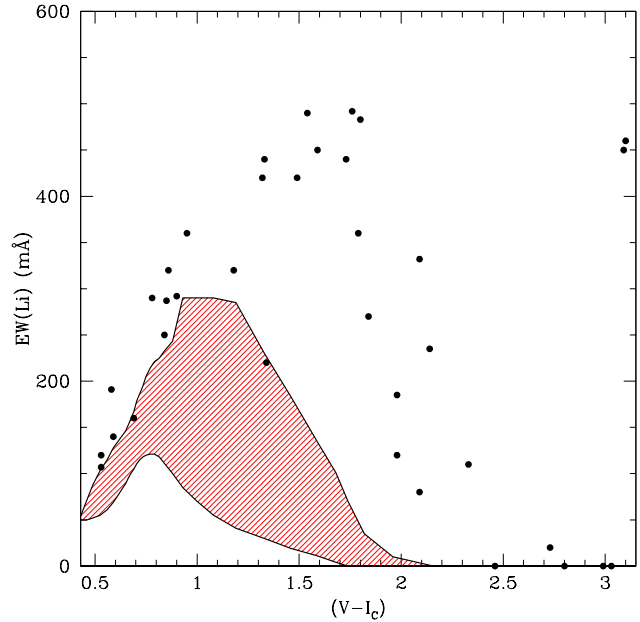}}
\caption{Distribution of Li equivalent width as a function of the $(V-I)_C$ for the
proposed members of the $\beta$~Pic Association shown in Table~\ref{table:betapic}. 
The hatched area shows the Li equivalent width range
observed for the Pleiades members.}
\label{fig:li_bpic}
\end{figure}

Figure~\ref{fig:li_bpic} shows the Li equivalent width as a function of color index
for all proposed members of Table~\ref{table:betapic}.
As seen in Figure~\ref{fig:li_bpic}, they lie above the upper envelope set
by the Pleiades members, as expected from the fact that
by definition the young sample contains only stars showing a considerable Li.
Nevertheless, it is worth noticing
that for the G-K dwarfs (where the Li-age correlation is stronger)
the observed equivalent widths are much higher than the Pleiades upper envelope 
indicating a much younger age
for these objects relative to the Pleiades.
We tested the convergence method including the Pleiades age sample without finding new members.

To obtain the Li abundances,  presented in Table~\ref{table:bpali}, we used our
modified version of the known LTE code made by Monique Spite, of Paris-Meudon Observatory. 
These Li abundances  were determined by comparing  the equivalent widths  of the
synthetic lines to the measured ones, on the observed spectra. The method is
very simple: given an equivalent width for a given line, we calculate its theoretical profile
for a given element abundance and its corresponding equivalent width. 
If the theoretical equivalent width is
not good, the process is retrieved for an interpolated abundance value, until the
difference between both values be acceptable. Our criterion is that this
difference be less than 0.2~m\AA.  In the computation of the synthetic profile we
take in account the 4 components of the $^7Li$ resonance line, being the
wavelengths and the oscillators strengths given by \citep{Andersen84}. 
The used stellar atmospheric models  are from Gustafsson and
collaborators \citep{Edvardsson93}. 
The effective temperatures used for the atmosphere models were determined from 
$(V-I)_C$ color index  using the \citep{Kenyon95} calibration.
The other atmosphere model parameters were fixed a priori: 
the  metallicity as \mbox{$[Fe/H]~=~0.1$} \citep{castilho05}  and \mbox{$\log~g~=~4.5$}, 
for stars of  luminosity class V, or \mbox{$\log~g~=~4.0$} for luminosity class IV. 
The  microturbulence velocity also was fixed as \mbox{2~km~s$^{-1}$}. 
The Li abundance is not very sensitive to  gravity and the main source
of errors is the inaccuracy of the equivalent widths (specially for fast
rotating stars) and  the effective temperatures.

\begin{figure}
  \resizebox{\hsize}{!}{\includegraphics{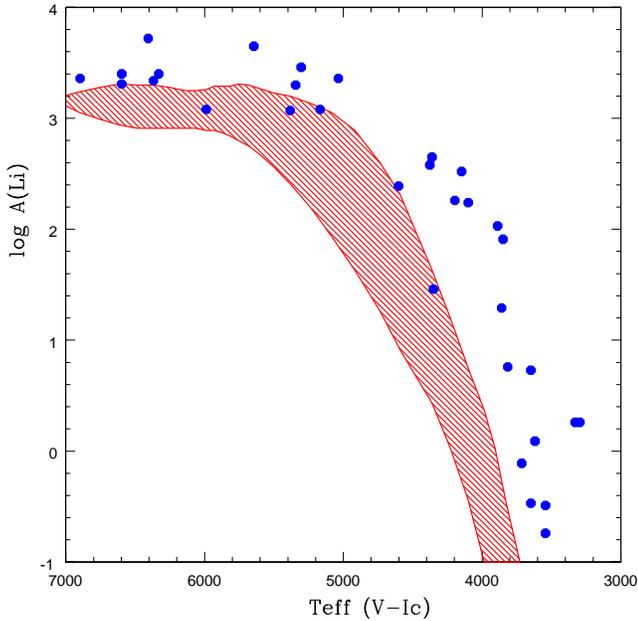}}
\caption{Distribution of Li abundances as a function of the effective temperature for the
proposed members of the $\beta$~Pic Association  (Table~\ref{table:bpali}). 
The hatched area shows the Li abundances with range observed for the Pleiades members}
\label{fig:t_bpic}
\end{figure}

\begin{table}
\caption{Li abundances, in the scale log A(H)=12,  for the BPA proposed members having EW$_{Li} > 0$
in Table~\ref{table:betapic}.
Column 1: Star identification; Column 2: Effective temperature, from $(V-I)_C$; Column
3: Equivalent width of the resonance lithium line in m\AA; Column 4: log of Li abundance.
}
\label{table:bpali}
\centering
\begin{tabular}{llrr}
\hline
~IDENT & T$_{\rm eff}$ & EW & A(Li) \\
~~~(1) & ~(2) & (3) & (4)~  \\

\hline
\hline

HD~203     &    6897. &   87   & 3.36 \\
Hip~10679  &    5988. &  160   & 3.08 \\
HD~14082  &     6368. &  140   & 3.34 \\
BD+30~397B  &   3544. &  110   &-0.49 \\
AG~Tri    &     4351. &  220   & 1.46 \\
BD+05~378  &    4100. &  450   & 2.24 \\
GJ~3305	  &     3715. &  120   &-0.11 \\
V1005~Ori  &    3816. &  270   & 0.76 \\
CD-57~1054  &   3860. &  360   & 1.29 \\
HD~35850  &     6406. &  191   & 3.72 \\ 
AO~Men    &     4377. &  420   & 2.58 \\
HD~139084B  &    3310. &  460   & 0.26 \\
V343~Nor  &     5168. &  292   & 3.08 \\
V824~Ara  &     5383. &  250   & 3.07 \\
HD~155555C  &    3413. &   20   &-1.76 \\
CD-54~7336  &   5036. &  360   & 3.36 \\
HD~161460  &    5304. &  320   & 3.46 \\  
HD~164249  &    6597. &  107   & 3.31 \\
HD~164249B  &   3544. &   70   &-0.74 \\
V4046~Sgr  &    4361. &  440   & 2.65 \\
GSC~7396-0759  &   3619. &  200   & 0.09 \\
HD~168210  &    5645. &  290   & 3.65 \\
CD-64~1208  &   4148. &  490   & 2.52 \\
TYC~9073-0762  &   3649. &  332   & 0.73 \\
TYC~7408-0054  &   3889. &  492   & 2.03 \\
PZ~Tel     &    5344. &  287   & 3.30 \\
TYC~6872-1011  &   3850. &  483   & 1.91 \\
CD-26~13904  &  4602. &  320   & 2.39 \\
HD~181327  &    6597. &  120   & 3.40 \\
AU~Mic       &  3649. &  80& -0.47\\
HD~199143  &    6330. &  150   & 3.40 \\  
AZ~Cap     &    4197. &  420   & 2.26 \\
CP-72~2713  &   3921. &  440   & 1.84 \\
TX~PsA 	   &    3316. &  450   & 0.26 \\
BD-13~6424  &   3715. &  185   & 0.20 \\

\hline
\end{tabular}

\end{table}

A further evidence for the youth of the BPA members is given by
the projected rotational velocities ($V\sin(i)$).
They  span over a large range from $\sim5$ up to $\sim150$~km~s$^{-1}$ as typically
observed in young clusters of similar age such as
IC~2391 and IC~2602 \citep{stauffer97} and the slightly older $\alpha$~Per \citep{Bouvier-1997}.
This spread is also in good agreement with the
theoretical angular momentum evolution tracks  and are the result
of different time-scales for disk-unlocking which is probably related to the
disk life-time itself \citep[e.g.][]{allain97}.

\subsection{Comments about some proposed members}

AT~Mic and AU~Mic were already classified as young stars by \citet{eggen1968} 
whereas their connection with $\beta$~Pic was first proposed by \citet{barrado99}.  
Errors in the Hipparcos distances for the AT Mic pair  are very high and their proper motions
have differences larger than usual between Hipparcos, Tycho-2 and UCAC2. 
The pair has an orbital solution with a period of 209~yr 
in the Sixth Catalog of Orbits of Visual Binary Stars (Hartkopf \& Mason, htpp://ad.usno.navy.mil/wds/orb6.html), 
thus catalogs with larger time base may have the proper motions contaminated by the orbital motion.
Therefore, we prefer to use the proper motions of Hipparcos, as they better reflect 
the center of mass motion.  
Our kinematical distance is close to that of the companion AU~Mic.
The weak Li line of AU~Mic was first suspected by \citet{delareza81} in an attempt to explain the odd
Li-rich red dwarf V1005~Ori (Hip~23200; Gl~182) by means of Li production by spallation reactions. 
They discarded  this possibility and proposed that V1005~Ori may be a member of a very young kinematical group.
These stars may now have their oddities explained as proposed members of the young BPA.

WW~PsA and TX~PsA form a low mass wide visual binary, 
the first one being  slightly brighter and hotter.
TX~PsA has a high Li equivalent width (the point around $(V-I)_C\sim3$ in Figure~\ref{fig:li_bpic}). 
In contrast, WW~Psa has the Li fully depleted. 
\citet{song02} interpreted the pair in the context of Li depletion boundary and concluded
that it can give a strong observational constraint to age determinations using the Li depletion boundary. 
Our kinematical distances put the pair closer to the Sun than the
Hipparcos parallax which has a large uncertainty. 
If our distance is correct, the stars could be fainter,
and this should be considered in the discussion of the Li age determinations.
The faint companion of HD~139084 is very similar to TX~PsA (the spectral type may be somewhat later)
and it can help in this context. The $(V-I)_C$ given by \citet{song03}
seems too blue; from 2Mass, the two stars have similar (V-K),  and we use in the paper the TX~PsA color for it.

HD~199143 (Hip~103311) and AZ Cap (HD~358623) were recently proposed by  
\cite{kaisler04} as members of the BPA. HD~199143 is a fast rotator
and its radial velocity has a large spread. 
If we had used the value given by \citet{ancker00}, 
the results yielded by the convergence method would support the BPA membership for both HD~199143 and AZ Cap.
Our results challenge the existence of the proposed Capricornius moving group \citep{ancker00},
which is more likely  a sub-sample of the BPA.

$\eta$ Tel is a very fast rotator. The $V\sin(i)$ in Table~\ref{table:betapic} was
estimated by \citet{gray87}. Thus its radial velocity is unreliable and
we used 0~km~s$^{-1}$ as proposed by \citet{song03}.
This value is very close to the radial velocity of its wide companion HD~181327. 
It has a proposed brown dwarf companion at 4.1$\arcsec$ \citep{Lowrance}.

There are 5 double line spectroscopic binaries (SB2) in the BPA:

Hip~23418 was found by \citet{delfosse} to be a SB2 with a period near 12 days
and mass ratio of 0.57. In Table~\ref{table:betapic} we used their systemic velocity. 
The star is also a very close visual double detected by Hipparcos, the secondary star
being one magnitude fainter. In Table~\ref{table:betapic} we try to give
the magnitude and color of the primary for this triple system.
 
HD~35850 was found by \citet{Nord04}  to be a SB2, who give a mass ratio of 0.72,
the systemic velocity, but no rotational velocity. 
For $V\sin(i)$ we use that determined by \citet{wichmann03}.
The star is considered an outlier by \citet{song03} since they did not know about
the radial velocity variations. 
It is also a variable star with a period near one day \citep{budding, wright}.

HD~155555 (V824~Ara) is actually a multiple system that includes a well studied SB2 (G7+K0IV)
\citep{strass} and a fainter optical companion, proposed as member by \cite{zuckermanSong04}.
We observed also this fainter companion, a M4Ve star with no detectable Li line, and our radial velocity
is consistent with the value obtained by \cite{pasquini91} (3.5~km~s$^{-1}$).
In Table~\ref{table:betapic} we used for its proper motions and parallax the data of the brighter component.

V4046~Sgr is not in the SACY sample, since it is not in 1RXS, but we included it
to test if it could belong to some association as it is an isolated classical TTS. 
It is also a SB2 with a circumbinary disk \citep{stempels04, quast00}. 
The distance obtained by \cite{quast00}, $83\pm8$~pc, evolutionary model dependent, 
agrees well with our kinematical one of 72~pc.
The age found by \cite{stempels04} also agrees well with the age of the BPA.
GSC 7396-0759, a nearby young red
star and optical counterpart of the ROSAT source 1RXS~J181422.6-324622, 
could be a distant (2.8\arcmin) physical companion of
the V4046~Sgr system, based on its radial velocity, Li equivalent  width and magnitudes.
Using for it the V4046~Sgr Tycho-2
proper motions, the star agrees well with a BPA membership.
However, the shape of its CCF suggests it may be a SB2 and, in this case,
it could not be a BPA member.

HD~161460 is a SB2 (K0IV + K1IV), observed only twice, one near conjunction and in the other spectrum
the lines are not very well resolved, but both Li lines are strong.
The radial velocity is reliable and the mass relation is near 0.5, implying a later spectral type for
the secondary star.

BD+05 378 is a single line spectroscopic binary detected by \cite{song03} whose estimated
systemic radial velocity is used here.
We observed it only once, after its proposed inclusion in the BPA,
and the other spectroscopic information in Table~\ref{table:betapic}
comes from our spectrum.

Thus, there are 6 known spectroscopic binaries, 6 close visual binaries (separation less than 200~AU)
and 10 wide visual binaries
(see Table~\ref{table:betapic} and \citet{zuckermanSong04}) in BPA over 41 (or 43) proposed members.

\section{Summary and conclusions}

In the present paper we present the SACY, a high-resolution optical spectroscopic survey
aiming to look for undiscovered nearby associations among the optical counterparts of
the ROSAT X-ray sources in the Southern Hemisphere.
From an initial list containing 9574 1RXS sources in the Southern Hemisphere, 
we find 1953 of them having
counterparts with $B-V\geq0.6$ in the Tycho-2  or Hipparcos Catalogues.
During the last seven years, 1511 objects were observed at least once.
Published information for 115 objects was used to complete our measurements.
SACY is $\sim80\%$ complete in the Southern Hemisphere.

The main results of this paper are summarized below.

\begin{itemize}

\item {\it Radial velocities and $V\sin(i)$}. Accurate radial velocities ($\sigma(V_r)\la0.3$~km~s$^{-1}$) 
have been derived for all observed sample using FEROS and/or coud\'e observations. 
Good quality $V\sin(i)$ were derived from FEROS observations. 
Both quantities, radial velocities and $V\sin(i)$, were compared to recent measurements 
obtained with CORAVEL \citep{Nord04} showing a very good agreement.

\item {\it Spectral classification}. Spectral type classification was carried out 
using a combination of three different methods: 
i) comparison with the \cite{montes97} spectral library, 
ii) calibrations of \cite{tripicchio97, tripicchio99} for the Na\,I~D and K\,I 7699~\AA\ lines and, 
iii) the method developed by \cite{Torres1998} in the $H_\alpha$ region.

\item {\it Young stars}. Based on our spectroscopic observations Li equivalent width 
could be measured for the first time for most of the 1256 dwarfs observed in the survey.
Using also the information  for 115 stars from the literature and  
based on the Li criterion we consider, 565, 282 and 524 dwarf stars as respectively younger, 
 having a similar age to or being older than the Pleiades.

\end{itemize}

Since parallaxes exist only for a small part of the sample, we developed
a kinematical convergence method to overcome this problem allowing us to derive
dynamic parallaxes.
As an example, this method is
applied to an initial concentration of young stars at U,V,W=-10,-16,-9~km~s$^{-1}$
which is very close to the proposed mean velocities for BPA by \citet{zuckermanSong04}.

\begin{itemize}

\item {\it Defining the BPA in (UVW)}. According to Table~\ref{table:betapic}, the association
has a mean (UVW) of ($-10.1\pm2.1$~km~s$^{-1}$, $-15.9\pm0.8$~km~s$^{-1}$, $-9.1\pm0.9$~km~s$^{-1}$).

\item {\it New members of the BPA}. Our definition of the BPA is close to that of
\cite{zuckermanSong04}. Using our convergence method and data from the literature we verify that  most of their
northern and early-type members fit well into our definition of the BPA.
According to our results we propose  13 additional members, 
including the well known active stars V1005~Ori and V4046~Sgr. 
This latter is not a "SACY star" and was formally considered as typical isolated TTS.
Our results suggest that V4046~Sgr belongs indeed to the BPA, being one
of the oldest cTTS known up to now.  We propose a wide companion red dwarf star for it.
We find also a faint companion for HD~164249. We propose a list of 41 {\it bona-fide} BPA members.

\item {\it Observational isochrone of the BPA}. The 10~Myr isochrone of \cite{siess00} fits part of the
temperature range of the members of the BPA. We then reversed the process and used the color-magnitude
diagram of the BPA to define an {\it ad-hoc} observational isochrone as done by \cite{zuckerman01}.

\item {\it Indication for expansion}. We find  a positive correlation 
between the X-direction and the U-component of
the space velocity vector for young stars in the SACY sample.  
This behavior is most clearly seen in the case of the BPA.
Analysis of this expansion and its implications will be discussed in a future paper.

\item {\it Membership probability}. We have developed a membership assessment procedure
based on a predictive classification scheme to be used with the convergence method. Both
the value of the $F$ and membership probability agree well.

\end{itemize}

\begin{acknowledgements} C.A.O.Torres was an ESO Visitor Scientist program
and thanks CNPq, Brazilian Agency, for the grant 200256/02.0.
L. da Silva thanks also the CNPq for the grant 30137/86-7.
We thank the Centre de Donn\'{e}es Astronomiques de Strasbourg (CDS) and 
NASA for the use of their electronic facilities, specially SIMBAD and  ADS.
This research made use of the Washington Double Star Catalog maintained at the U. S. Naval Observatory.
\end{acknowledgements}

\bibliographystyle{aa}

\end{document}